\documentclass[aps, prl, reprint, superscriptaddress, amsmath, amssymb, preprintnumbers, nobibnotes, nofootinbib]{revtex4-2}
\usepackage{fancyhdr}
\usepackage{graphicx}
\newcommand{\plotfolder}{}

\usepackage{dcolumn}
\usepackage{bm}
\usepackage{hyperref}
\usepackage{cleveref}

\usepackage[mathlines]{lineno}

\usepackage{xcolor}
\usepackage{import}
\usepackage{lipsum}
\usepackage{multirow}

\usepackage{siunitx}
\usepackage{booktabs}
\usepackage[version=4]{mhchem}

\begin{document}
{
\title{Search for dark matter particle interactions in an extended nuclear recoil  energy window with the LUX-ZEPLIN (LZ) experiment}



\author{D.S.~Akerib}
\affiliation{SLAC National Accelerator Laboratory, Menlo Park, CA 94025-7015, USA}
\affiliation{Kavli Institute for Particle Astrophysics and Cosmology, Stanford University, Stanford, CA  94305-4085 USA}

\author{A.K.~Al Musalhi}
\affiliation{University College London (UCL), Department of Physics and Astronomy, London WC1E 6BT, UK}

\author{B.J.~Almquist}
\affiliation{Brown University, Department of Physics, Providence, RI 02912-9037, USA}

\author{C.S.~Amarasinghe}
\affiliation{University of California, Santa Barbara, Department of Physics, Santa Barbara, CA 93106-9530, USA}

\author{A.~Ames}
\affiliation{SLAC National Accelerator Laboratory, Menlo Park, CA 94025-7015, USA}
\affiliation{Kavli Institute for Particle Astrophysics and Cosmology, Stanford University, Stanford, CA  94305-4085 USA}

\author{T.~Anderson}
\affiliation{Lawrence Berkeley National Laboratory (LBNL), Berkeley, CA 94720-8099, USA}

\author{N.~Angelides}
\affiliation{University of Zurich, Department of Physics, 8057 Zurich, Switzerland}

\author{H.M.~Ara\'{u}jo}
\affiliation{Imperial College London, Physics Department, Blackett Laboratory, London SW7 2AZ, UK}
\affiliation{STFC Rutherford Appleton Laboratory (RAL), Didcot, OX11 0QX, UK}

\author{J.E.~Armstrong}
\affiliation{University of Maryland, Department of Physics, College Park, MD 20742-4111, USA}

\author{M.~Arthurs}
\affiliation{SLAC National Accelerator Laboratory, Menlo Park, CA 94025-7015, USA}
\affiliation{Kavli Institute for Particle Astrophysics and Cosmology, Stanford University, Stanford, CA  94305-4085 USA}

% ...

\collaboration{The LZ Collaboration}

\begin{abstract}
We report on a search for dark matter particles interacting with xenon nuclei in an exposure of 2.84~tonne-years with the LUX-ZEPLIN (LZ) experiment. 
An extended nuclear recoil energy window up to approximately 270~keV enables searches for effective field theory and inelastic models of dark matter where high-energy recoils account for a larger fraction of the predicted recoil spectrum compared to the nominal spin independent interaction.
We observe one event with characteristics consistent with a nuclear recoil of $248\pm23\,\mathrm{(stat)}\pm23\,\mathrm{(sys)}$~keV, in a region where the known background expectation is low. 
A profile likelihood ratio test finds tension with the background-only hypothesis at a global significance of 2.6$\sigma$ when accounting for look-elsewhere effects, with a maximum local significance of 3.4$\sigma$ across the models tested. 
We describe the analysis and event of interest, the background model used in the statistical inference, and detail several of the rare background topologies considered.  
\end{abstract}

\maketitle
}
\thispagestyle{fancy}

\paragraph{\label{sec:introduction}Introduction --}

Astrophysical and cosmological observations have provided compelling evidence for the existence of dark matter (DM) that would constitute most of the matter content of the Universe~\cite{1937ApJ,1970ApJ...159..379R,Clowe_2006,Planck2018VI}.
Weakly Interacting Massive Particles (WIMPs) are among the best-motivated candidates for dark matter, arising naturally in many theories beyond the Standard Model.
Numerous direct-detection experiments have searched for interactions between dark matter particles and ordinary matter, with experiments such as LUX-ZEPLIN (LZ) placing stringent constraints on many leading theoretical models~\cite{LZ:SR3_WS2024, XenonnT:WIMP-SI-2025,PandaX4T:WIMP-2025,DEAP3600:2026_WIMP,DarkSide-50:2018_532days,SuperCDMS:2018_WIMP,EDELWEISS:2019_WIMP,CRESST:2019_DM,PICO:2019_DM}.

WIMP searches for particles with mass  $\gtrsim10$~GeV/c$^2$ typically focus on spin-independent (SI) and spin-dependent~(SD) interactions generating $\lesssim100$~keV nuclear recoils~\cite{Lewin:1995rx, Goodman:1984dc, Griest:1988ma}.
Any momentum dependence in these interactions is assumed to be suppressed as the inverse of the momentum transfer is much larger than the size of the nucleon. 
Allowing for momentum-dependent WIMP-nucleon couplings can result in predicted recoil spectra that differ from the SI and SD cases, including non-negligible contributions at higher recoil energies~\cite{Chang:2009yt,Feldstein:2009tr}. 
In this Letter, we report the latest results from the LZ experiment in a nuclear recoil energy range up to 270~keV.

\paragraph{Theory --\label{sec:theory}}
Effective field theories, such as those developed by Fan et al.~\cite{Fan_2010} and Fitzpatrick et al.~\cite{Fitzpatrick:EFT}, provide a more general framework for DM scattering than the usual SI and SD treatment.
We use the non-relativistic effective field theory (NREFT) developed in Ref.~\cite{Fitzpatrick:EFT}, in which the WIMP-nucleon interaction is treated as a four-particle contact interaction
\begin{equation}
    \mathcal{L}_\text{int} = \mathcal{O}\chi\bar{\chi}\bar{N}N,
\end{equation}
where $\chi$ and $N$ are non-relativistic fields for the WIMP and target nucleon, respectively.
Enforcing momentum conservation and Galilean invariance \cite{Fitzpatrick:EFT}  reduces the operators to a basis of four Hermitian quantities
\begin{equation}
i \frac{\vec{q}}{m_N}, \quad \vec{v}^\perp \equiv \vec{v} + \frac{\vec{q}}{2\mu}, \quad \vec{S}_{\chi}, \quad \vec{S}_{N},
\label{eq:basis}
\end{equation}
where $\vec{q}$ is the momentum transferred from the WIMP to the nucleon, $m_N$ is the nucleon mass, $\mu$ is the reduced mass of the WIMP-nucleon system, $\vec{v}$ is the relative velocity between the WIMP and the nucleon, $\vec{v}^\perp$ is the component of the relative velocity  perpendicular to the momentum transfer, $\vec{S}_{\chi}$ is the spin of the WIMP, and $\vec{S}_{N}$ is the spin of the relevant nucleon. 
Linear combinations of these Hermitian quantities up to 2$^{\rm nd}$-order in $\vec{q}$ give a total of fifteen independent and dimensionless NREFT operators, denoted as $\mathcal{O}_i$. The operators can be mapped onto fully relativistic covariant Lagrangians, the nonrelativistic reductions of which we denote $\mathcal{L}_i$, allowing for more complex interactions to be studied~\cite{PandaX:2023_dm_luminescence,LZ:SR1_NREFT_Lagrangian_2024,Anand:MathematicaEFT}.
Equation~\ref{eq:basis} can also be modified to describe inelastic scatters where the WIMP transitions into a more massive state during the interaction, another mechanism that can lead to a suppression of the low energy component of the recoil spectrum~\cite{Barello_2014}. 
Inelastic models arise outside of NREFT, and they can generate interaction rates with significant annual modulation peaking around June 2~\cite{smith_2021:inelasticDM,Tucker-Smith:inelasticDM,Bramante2016:inelasticDM,Graham:Higgsino_2025,Han_1997}.

\begin{figure}
    \includegraphics[width=0.95\columnwidth]{\plotfolder 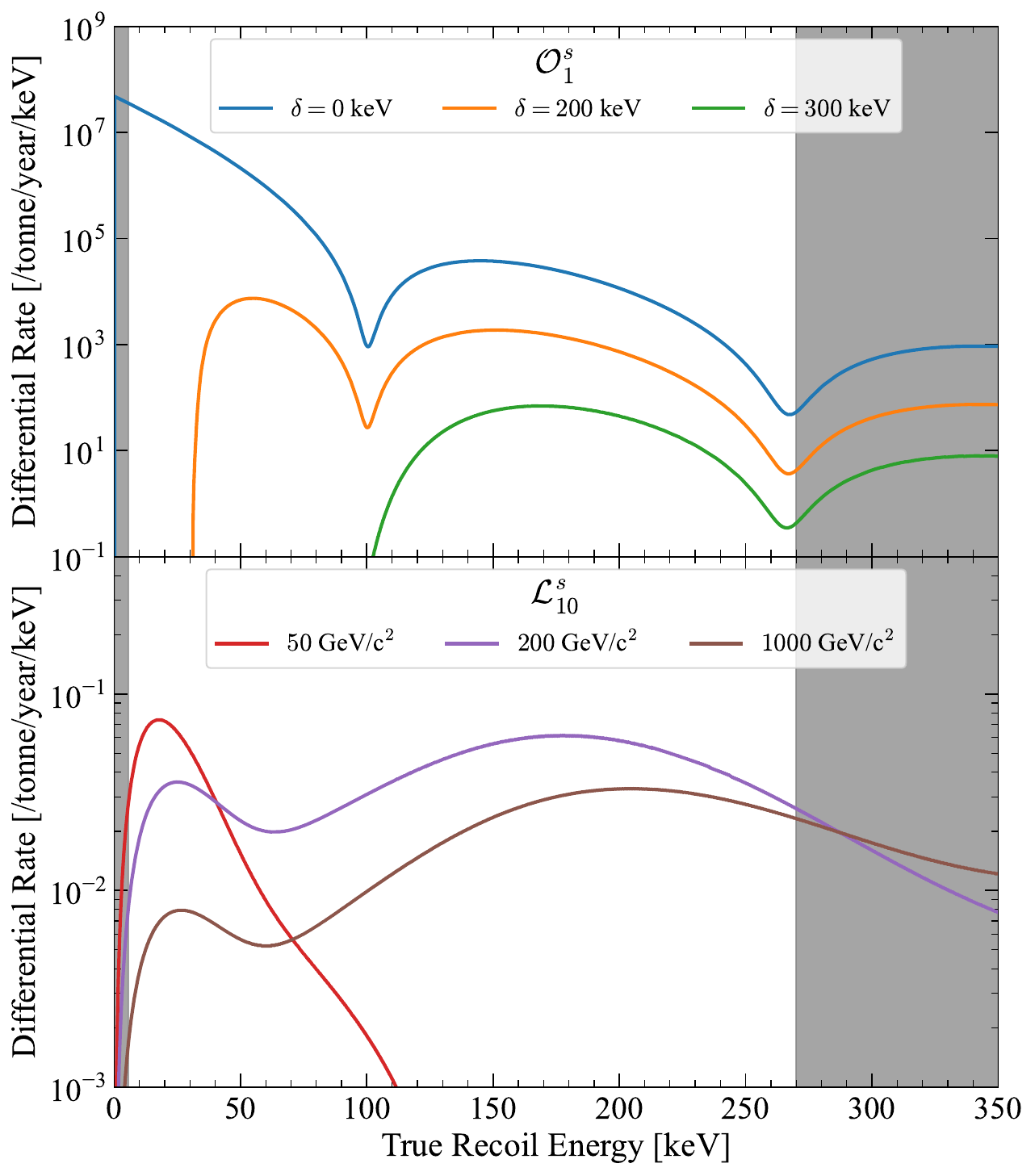}
    \caption{Nuclear recoil energy spectra for the two interactions chosen as examples in this Letter.
    \textbf{Top:} Interactions via the inelastic $\mathcal{O}^{s}_{1}$ with splitting between WIMP energy levels ($\delta$) of 0~keV (equivalent to a purely elastic spin-independent interaction), 200~keV, and 300~keV for an impinging WIMP of mass 1000~GeV/c$^2$. 
    \textbf{Bottom:} Elastic interactions of a 50, 200, and 1000~GeV/c$^2$ WIMP via $\mathcal{L}^{s}_{10}$.
    The shaded gray regions, $<$5.4~keV and $>$269.9~keV, indicate where the detection efficiency is below 50\% after all data analysis criteria.
    The coupling strength of each interaction is assumed to be unity 
    (\emph{i.e.}, $c_i^s = 1/m_\nu^2$ as in Eq. 21 and $d_i^s = 1/m_\nu^2$ as in Eq. 68 of Ref.~\cite{Anand:MathematicaEFT}, where $m_\nu = 246.2$~GeV/c$^2$ is the Higgs vacuum expectation value.). 
    }
    \label{fig:signal_recoil_spectra}
\end{figure}

Here, we perform a search for high-energy nuclear recoils arising from $\mathcal{L}_{1}-\mathcal{L}_{20}$~\cite{Anand:MathematicaEFT}, and for inelastic interactions from $\mathcal{O}_1$ and $\mathcal{O}_4$
with superscripts ($s$) and ($v$) referring to the isoscalar and isovector basis, respectively.
The inelastic operators are selected because they are extensions of SI and SD models.
We use two interactions as illustrative examples, shown in~\autoref{fig:signal_recoil_spectra}. 
The first is the inelastic spin-independent scatter, $\mathcal{O}^{s}_{1}$, which resembles a  Higgsino model~\cite{Graham:Higgsino_2025}.
The second is $\mathcal{L}^{s}_{10}$, which describes a magnetic moment interaction, chosen as it results in a broad and double-peaked spectrum in our region of interest (ROI). 
Predicted xenon nuclear recoil spectra for all dark matter models are calculated with WimPyDD~\cite{wimpydd:Jeong_2022} using one-body nuclear density matrices developed for DMFormfactor-v6~\cite{Anand:MathematicaEFT}, with modifications as described in Ref.~\cite{LZ:SR1_NREFT_2023}, and the Standard Halo Model with parameters recommended in Ref.~\cite{DM_parameters:BAXTER2021_Conventions}. 
Results for all of the probed interactions are in the \textcolor{black}{Data Release}.

\paragraph{LZ Detector and Calibrations --}
The heart of the LZ experiment is a low-background dual-phase time projection chamber (TPC) containing 7 tonnes of active liquid xenon (LXe)~\cite{LZ:Experiment_2020, LZ:TDR_2017}.
The TPC is equipped with arrays of photomultiplier tubes (PMTs) at its top and bottom to detect optical signals.
Cathode, gate, and anode electrodes create a set of electric fields that drift ionization electrons to the liquid surface and extract them into the gas phase.
The liquid-gas interface lies between the gate and anode at the top of the detector, and the cathode defines the bottom of the active region. A bottom electrode sits below the cathode to shield the lower PMT array.

The TPC  is surrounded by a LXe Skin veto detector, consisting of 2~tonnes of LXe instrumented for scintillation readout whose main purpose is to identify $\gamma$-rays  entering or exiting the TPC.
The TPC and Skin are contained within a vacuum-insulated cryostat, which is in turn surrounded by the Outer Detector (OD) veto. The OD is 17~tonnes of gadolinium-loaded liquid scintillator (0.1\% by mass) and 229~tonnes of ultra-pure water, read out by PMTs in the water space, and it detects neutrons, $\gamma$-rays, and muons.
The LZ detectors are described in more detail in Ref.~\cite{LZ:Experiment_2020}.

Energy depositions in the TPC typically produce two signals: prompt scintillation light (S1), produced at the location of the interaction; and delayed electroluminescence light (S2), produced when extracted electrons are accelerated through the gaseous phase by the applied electric field.
The S2 light hit pattern in the top PMT array enables position reconstruction in the ($x,y$) plane, and the time interval between S1 and S2 pulses provides the depth, $z$. S1 and S2 signal sizes are measured in photons detected (phd) and vary with the position of the interaction vertex. Position-corrected values, denoted S1$c$ and S2$c$, allow for scale factors to correlate the signals to the original number of photons and electrons produced.
The photon gain factor $g_1=0.110\pm0.002$~phd/photon and the electron gain factor $g_2=34.5\pm1.1$~phd/electron are determined using monoenergetic peaks from background and calibration sources.

\begin{figure}
    \centering
    \includegraphics[width=0.95\linewidth]{\plotfolder 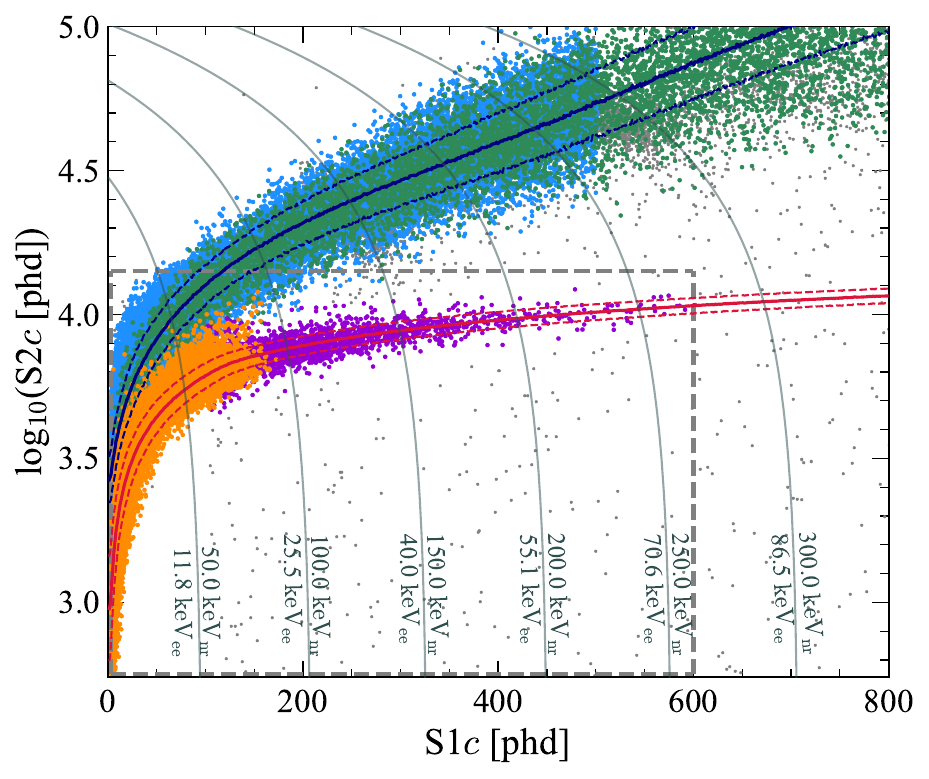}
    \caption{
    Calibration events in \{S1$c$,~log$_{10}$(S2$c$)\} space from tritium and $^{14}$C (light blue), $^{212}$Pb (green), D-D neutrons (orange), and AmBe neutrons (purple). 
    The mean (solid) and 10\%-90\% percentiles (dashed) of the ER (dark blue) and NR (red) bands are given by the calibrated NEST model, taking flat energy spectra and evaluating along lines of fixed S$1c$. The  solid grey lines show contours of constant energy in \{S1$c$, log$_{10}$(S2$c$)\} space. The dashed grey line indicates the WS~ROI.
    Tritium and ${}^{14}$C data are cut off above S1$c=500$~phd to avoid ${}^{133}$Xe contamination from activation due to a preceding neutron calibration. 
    To increase statistics, the calibration datasets include events outside the fiducial volume used in the main WIMP search analysis, leading to some wall leakage. Grey points are outside of $4\sigma$ from the mean of the ER and NR bands. These points are not used in deriving the NEST model as they are likely from interactions near the TPC walls.  
    }
    \label{fig:calibrations}
\end{figure}

The S1 to S2 ratio provides discrimination power between electron recoil (ER) events, produced mainly by $\gamma$-ray and $\beta$-decay interactions, and nuclear recoil (NR) events, expected from dark matter interactions as well as neutrons and coherent scattering of neutrinos -- as illustrated in~\autoref{fig:calibrations} with calibration datasets. 
The ER response is measured using \textit{in-situ} $\beta$-decays of tritium and $^{14}$C, injected as methane, and $^{212}$Pb from $^{220}$Rn injections.
The NR response is measured using neutrons from Deuterium-Deuterium (D-D) fusion (0--74~keV NRs) and from AmBe ($\alpha$,n) reactions (0--330~keV NRs)~\cite{LZ:Calibrations_2024}. 
We define the ``ER band'' and ``NR band'' as the distributions in \{S1$c$, log$_{10}$(S2$c$)\} space populated by flat-in-energy ER and NR recoil spectra, shown by the solid and dashed lines in Fig.~\ref{fig:calibrations}. The detector and model response parameters from NEST v2.4.5 (Noble Element Simulation Technique)~\cite{ NEST:paper_2023, nest_v2_4_5} are tuned to describe the ER and NR calibration data. 
The best-fit NEST parameters are provided in the \textcolor{black}{Data Release}, with more detail of the model in the \textcolor{black}{Supplemental Material}.

\paragraph{Data Analysis --}
The 220~live days of data used in this analysis were collected between 27~March 2023 and 1~April 2024.
The detector conditions and a search through these data for WIMPs producing signals with S$1c$ in the range  $3-80$ phd ($>50$\% acceptance for nuclear recoils of 5.4 to 55 keV)  are described in Ref.~\cite{LZ:SR3_WS2024}, which this analysis takes as a starting point. 
Events are selected as signal candidates if they are classified as single scatters (SS), pass the data and analysis selections, and are within a signal region of interest (ROI) and a fiducial volume (FV). The  analysis selections are unchanged from Ref.~\cite{LZ:SR3_WS2024} with the exception of the  ROI, the FV, and the handling of sidebands.
Here, the WIMP search ROI (WS~ROI) is defined by S1$c$ in the range $3-600$~phd, S2 greater than $645$~phd, and S2$c$ in the range $10^{2.75} - 10^{4.15}$~phd.
The maximum S2$c$ is chosen to keep essentially all of the signal distribution while eliminating the vast majority of ER backgrounds.
The S1$c$ range is the same as used in the previous NREFT search with LZ~\cite{LZ:SR1_NREFT_2023}.
The NR signal efficiency averages 96\% between 14~keV and  250~keV (shown in \autoref{fig:efficiency} in the \textcolor{black}{Supplemental Material}).

\begin{figure}
    \centering
    \includegraphics[width=0.95\linewidth]{\plotfolder 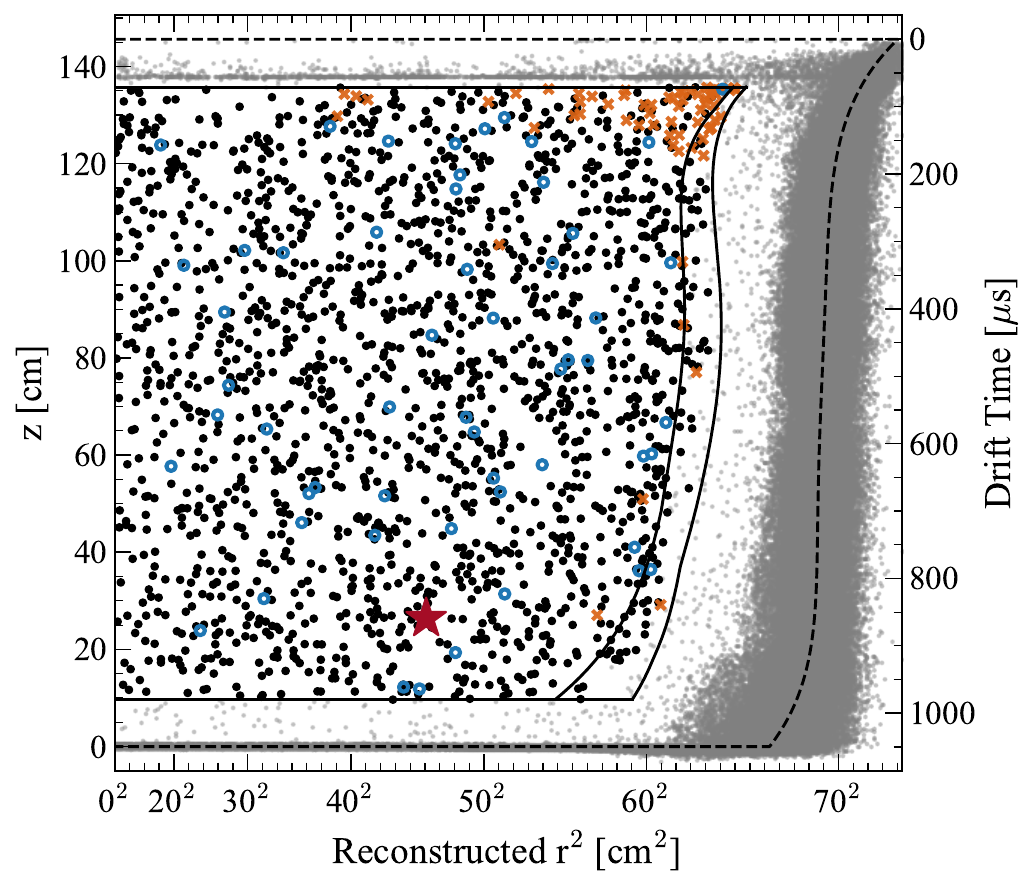}
    \caption{
    Data in the WS~ROI for the three analysis samples; \emph{science} (black), \emph{prompt veto} (orange crosses), and \emph{delayed veto} (blue circles).
    The event of interest is shown by the brown star near coordinate $(45.9^2\, \mathrm{cm}^2, 26.4\,\mathrm{cm})$.
    The gray points indicate data in the WS~ROI but outside the FV, with large populations of events at the cathode ($z=0$) and wall, as well as a population from interactions in the gas that reconstruct to a drift time of $\sim56$~$\mu$s.
    The dashed line shows the reconstructed wall, averaged over azimuth, indicating the active volume.
    The solid lines indicate the FV at the smallest and largest radial extents. The points in the \emph{delayed veto} sample come primarily from random coincidences of background pulses in the veto detectors with ER backgrounds in the TPC. 
    }
    \label{fig:roi_fv}
\end{figure}

The LXe mass contained within the FV used in this analysis is $4.71\pm0.08$~tonnes, 14.5\% smaller than in Ref.~\cite{LZ:SR3_WS2024}. 
Shown in~\autoref{fig:roi_fv}, the FV boundaries are chosen to limit backgrounds originating from the walls of the TPC due to detector radioactivity and plate-out of radon daughters, and from multiple-scintillation single-ionization (MSSI) events, discussed in more detail below.
As represented by the dashed line in~\autoref{fig:roi_fv}, the average reconstructed position of radioactive decays on the wall of the TPC is pushed inward due to electrostatic field effects, leading to a discrepancy between the true and reconstructed locations of the wall~\cite{LZ:SR1Backgrounds_2022, LZ:SR3_WS2024}. 
Events that occur on the wall can be further misreconstructed radially inwards due to finite position resolution. 
The radial boundary of the FV is a depth- and azimuth-dependent contour defined such that the wall backgrounds have an expected total count of less than $(1.0 \pm 0.5)\times10^{-4}$ events inside the FV.
The FV definition also enforces a minimum stand-off of 8.0 (6.0)~cm from the true (reconstructed) wall to the contour to reduce the MSSI contribution. 
The maximum distance of the FV contour to the true (reconstructed) TPC wall is 18.2 (10.4)~cm near the bottom of the TPC, and the mean distance is 10.7 (7.2)~cm.
The upper FV boundary is placed 12.8~cm below the gate electrode, and the lower boundary is placed 9.0~cm above the cathode.

\begin{figure}
    \centering
    \includegraphics[width=0.95\linewidth]{\plotfolder 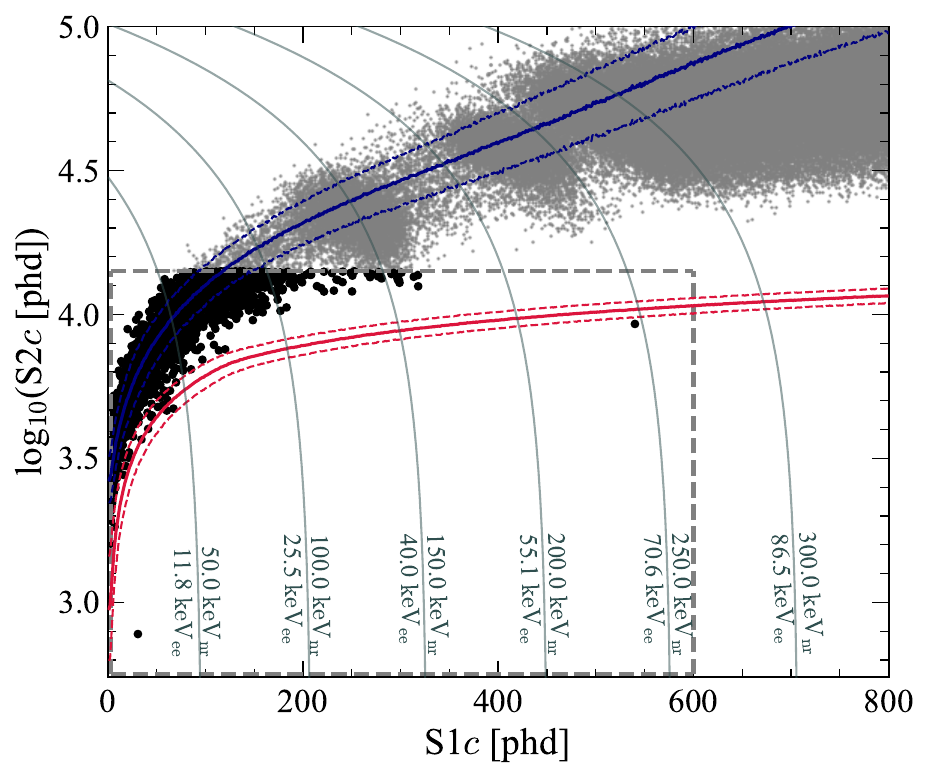}
    \caption{
    Black points show the WIMP search data from the full \emph{science} sample which pass all data selections and are not tagged by the veto detectors  in \{S1c, log$_{10}$(S2c)\} space.
    The dashed gray line indicates the WS~ROI, and the red and blue lines represent the NR and ER bands as in~\autoref{fig:calibrations}. The event of interest appears at (540.1~phd, $10^{3.98}$~phd). 
    The event below log$_{10}$(S2$c)=3$ is consistent with accidental background, as described in Ref.~\cite{LZ:SR3_WS2024}. The spectral features in the ER band above 25 keV$_\mathrm{ee}$ come predominantly from $^{125}$I and $^{133}$Xe decays and appear slightly below the beta-decay ER band due to increased recombination form Auger cascades, as described in Ref.~\cite{LZ:2025hud}.
    }
    \label{fig:finaldataset-S1-logS2}
\end{figure}

The data are divided into three samples according to the presence of veto signals. 
Events that have no signal in either veto system are classified as \emph{science} events, shown in~\autoref{fig:finaldataset-S1-logS2}.
Events with a signal of size $> 2.5$~phd in the Skin within $\pm 0.25$~$\mu$s of the TPC S1, or $> 4.5$~phd in the OD within $\pm 0.3$~$\mu$s, are classified as \emph{prompt veto}. 
Events with Skin or OD signals above 300 or 200 keV, respectively, that are within $600$~$\mu$s of the TPC S1 but not in the \emph{prompt veto} window are classified  as \emph{delayed veto}.
Simulations indicate that to a good approximation the signal and background components have the same \{S1$c$,~log$_{10}$(S2$c$)\} shape in each sample but have different rates.
The partition allows $\gamma$-rays and neutrons to be constrained more effectively as the rates of these backgrounds in the different samples are related by tagging efficiencies: the prompt window tags both $\gamma$-rays and neutrons, while the delayed window tags neutrons. 
Random coincidence with Skin and OD backgrounds cause 0.01\% (2.86\%) of unrelated TPC signals to be vetoed by prompt (delayed) pulses. More details on each sample can be found in the \textcolor{black}{Supplemental Material}.

An attempt to mitigate possible analyzer bias by injecting the dataset with artificial signal-like events, a process known as ``salting", was unsuccessful. Drawn from a distribution consisting of a sum of a WIMP SI-like exponential and a flat NR spectrum out to 250~$\pm$~25~keV, the salt events followed a NR model defined before the AmBe calibration was conducted and did not adequately cover the signal region at high energies.  
Despite the lack of coverage, the analysis selections and likelihood models were finalized before four salt events were revealed (in addition to the seven salt events previously removed as part of the analysis in Ref.~\cite{LZ:SR3_WS2024}). Three of the salt events fell in the WS~ROI with one just above, but all four fall outside the 90\% containment region for the calibrated NR band. The salt can be seen in the \textcolor{black}{Supplemental Material}. 
Bias mitigation in the region above 55~keV therefore relies primarily on the use of unchanged analysis selections from Ref.~\cite{LZ:SR3_WS2024}.
Ultimately, we consider the analysis described here to be a non-blind analysis.

After all selections are applied and salt is removed, one event in the \emph{science} sample stands out in the NR band, consistent with a $248\pm23\,\mathrm{(stat)}\pm23\,\mathrm{(sys)}$~keV elastic nuclear recoil (S1$c = 540.1$~phd, S2$c = 9268$~phd) and clearly visible in~\autoref{fig:finaldataset-S1-logS2}. The event was recorded at 21:22:39 UTC, 16 June 2023. The location of the event in \{S1$c$, log$_{10}$(S2$c$)\} is 1.5$\sigma$ below the median of the modeled NR band at that S1$c$, and it is 6.7$\sigma$ below the median of the modeled ER band.
The reconstructed position of the event is 26.4~cm above the cathode and 26.9 (23.4)~cm radially inward from the true (reconstructed) position of the TPC wall, illustrated in~\autoref{fig:roi_fv}. The S1 partitioning between the two PMT arrays is consistent with the reconstructed $z$ position of the vertex, and the S2 pulse shape is consistent with a point-like interaction at the reconstructed depth. Analysis of the S1 pulse shape does not allow for conclusive ER-NR discrimination, with more detail in the \textcolor{black}{Supplemental Material}.

All of the backgrounds described in Ref.~\cite{LZ:SR3_WS2024} are present in this analysis, with several additional ones considered. The dominant ER components are: $\beta$-decays of $^{214}$Pb and $^{212}$Pb in radon decay chains; long-lived radioisotopes $^{85}$Kr, $^{136}$Xe, and $^{124}$Xe; electron scattering from solar neutrinos; $\gamma$-rays from detector materials; residual tritium and $^{14}$C from calibration; and ${}^{125}$Xe and ${}^{127}$Xe produced by neutron capture during NR calibrations.  
The larger WS ROI introduces two additional ER sources not discussed in Ref.~\cite{LZ:SR3_WS2024}. First, WIMP search data-taking is periodically interrupted by injections of $^{83\mathrm{m}}$Kr ($t_{1/2}=1.83$\,hr), resuming approximately 13 half-lives after the conclusion of the injection. 
The residual rate of $^{83\mathrm{m}}$Kr in the \emph{science} sample is constrained using the observed decay rates in the intermediate period following the injection. Second, ${}^{125}$I is a daughter of ${}^{125}$Xe that decays via electron capture plus electron or $\gamma$-ray emission. 
The decay of  ${}^{125}$I has a half-life of 59.4 d, but it is efficiently removed by the purification system with an effective half-life in LZ of $t_{1/2}^{\mathrm{eff}} = 3.6 \pm 0.2$~d~\cite{LZ:2024wvs}. 
The rate of ${}^{125}$I is derived from measured ${}^{125}$Xe rates following neutron calibrations.

Neutrons and neutrinos are sources of NR background. 
Cosmic-ray-induced atmospheric neutrinos can produce NR in the WS ROI via coherent elastic neutrino--nucleus scattering (CE$\nu$NS).  
As in Ref.~\cite{LZ:SR3_WS2024}, the atmospheric neutrino recoil spectrum used in the background model is calculated using the formalism of Ref.~\cite{Billard:2013qya}, and flux normalizations and uncertainties follow the standard values recommended in Ref.~\cite{DM_parameters:BAXTER2021_Conventions}.
The shape of the neutron background is derived by simulation of ($\alpha$,n) activity in detector materials, as described in Ref.~\cite{LZ:SR3_WS2024}. 
The neutron rate is constrained in the analysis solely by the veto detectors, which have a total tagging efficiency for ($\alpha$,n) neutrons scattering in the TPC of 92$\pm$4\%.

Isolated S1 events, predominantly from interactions in charge-insensitive regions or pile-up of photoelectron backgrounds, and isolated S2 events, typically from spurious electron emission from the electrodes, can be coincidentally paired, causing ``accidental'' events which do not follow the typical ER band.
We use the same framework to model accidental backgrounds as described in Ref.~\cite{LZ:SR3_WS2024}, drawing random pairs from isolated S1 and S2 pulse populations.
The normalization is constrained using ``unphysical drift time'' events where the drift time is greater than that of interactions near the cathode.

The final background in the model is from MSSI events, or events in which a standard ER process deposits energy in two or more locations, with all but one being in a ``charge-dead'' region where no ionization is collected.
The S1s merge together in time while the S2 of all but one of the deposits is lost, creating an event that is more NR-like.
There are two main regions of zero charge collection that contribute to MSSI: the 13.75 cm deep layer of LXe in the reverse field region (RFR) below the cathode, where the charge is drifted towards the bottom shield grid; and LXe within a few mm of the TPC wall where inhomogeneous fields can drive ionization electrons into the wall where they are absorbed.
The dominant contributors to these topologies are $^{214}$Pb decays in the LXe and $\gamma$-rays originating from detector components (other contributors are discussed in the \textcolor{black}{Supplemental Material}).
The leading ER background from $^{214}$Pb is a $\beta$ decay to the ground state of $^{214}$Bi with no associated $\gamma$-ray emission.
The MSSI background comes from excited-state decays, where the $\beta$ particle deposits a small amount of energy in the TPC and an associated $\gamma$-ray interacts in a charge-dead region.
For $\gamma$-rays from detector components, MSSI are caused by double scatters where one of the interactions is in a charge-dead region.
For these two sources, wall MSSI have a high probability of being promptly vetoed in the Skin or OD as the outgoing $\gamma$-ray can deposit energy in those locations. 

Simulations of the detector radioactivity and $^{214}$Pb events are used to derive the MSSI model.  The rate of detector radioactivity in the simulation is normalized by radioassay results and validated by data as in Ref.~\cite{LZ:SR1Backgrounds_2022}. The $^{214}$Pb MSSI rate is constrained in the same way as the SS $^{214}$Pb component. The final model combines both sources of MSSI together. 
A \emph{prompt veto} tagging efficiency of $94\pm2$\% for MSSI in the WS ROI is derived from the same simulation, again applied to both sources. An additional uncertainty arises from the modeling of the geometry of the charge-dead regions along the wall. 
The simulation models the charge-dead regions as a scalloped pattern extending up to 3~mm from the physical wall due to field-shaping rings in the TPC wall, based on analysis of wall populations and electrostatic modeling of the electric field~\cite{dey2025a}. 
The charge-dead volume in the model comprises 0.3\% of the total active volume. To reflect the uncertainty associated with the geometry of the charge loss regions, we include a 100\% uncertainty on the modeled MSSI rate. As the wall MSSI rate is proportional to the volume of the charge dead region in this regime, the uncertainty encompasses a charge dead region covering up to $0.6$\% of the active volume.

Probability density functions in \{S1$c$,~log$_{10}$(S2$c$)\} space for each of the background components and signal models are created using the BACCARAT package based on GEANT4~\cite{LZ:simulations_2021, GEANT4:2002zbu, ALLISON2016, Allison:2006ve}, with the tuned NEST v2.4.5~\cite{NEST:paper_2023} model providing the detector response. 
Exceptions are $^{83\text{m}}$Kr, which is taken directly from calibration data, and the accidentals model as noted above. 
The expected numbers of events in the WS~ROI for each background are given in~\autoref{tab:backgrounds} for the \emph{science} sample.

Simultaneous fits of the \emph{science}, \emph{prompt veto}, and \emph{delayed veto} samples 
are performed in the  \{S1$c$, $\log_{10}(\mathrm{S}2c)$\} plane using an unbinned, extended maximum likelihood. 
The normalization of each background model component is included as a nuisance parameter in the fit; uncertainties on the normalization are included as Gaussian constraint terms except for neutrons from detector materials, which
are constrained solely by the efficiencies associated with the \emph{prompt veto} and \emph{delayed veto} samples.
The MSSI, $^{127}$Xe, and $\gamma$-rays from detector materials are additionally constrained by the veto sidebands.

\begin{table}[th]
    \centering
    \caption{The expected and fitted numbers of events in the \emph{science} sample from all model components in the 220~d~$\times$~4.71~t exposure after all selections.
    The middle column contains the predicted number of events with uncertainties as described in the text. Detector NRs are neutrons from detector radioactivity, with the constraint provided solely from the veto sideband. 
    These uncertainties are used as constraints in a combined fit of the background plus signal model to the data in the \{S1$c$, $\log_{10}(\mathrm{S}2c)$\} plane.  
    The result of the fit for the 1000~GeV/c$^2$ $\mathcal{L}^s_{10}$ WIMP model is shown in the right column, and the post-fit uncertainty on the total counts includes correlations among the fitted parameters.
    }
    \begin{tabular}{l
                r@{$\,\pm\,$}l
                r@{$\,\pm\,$}l}
     \hline 
     \hline 
    Source & \multicolumn{2}{c}{Expected Events} & \multicolumn{2}{c}{Fit Result} \\\hline 
    Internal $\beta$ decays         & 1341   & 160           & 1340   & 43 \\
    $\nu$ ER                        & 140.6  & 8.4           & 140.3  & 8.4 \\
    $^{136}$Xe                      & 110.0  & 16.5          & 114.2  & 16.1 \\
    CH$_3$T \& $^{14}$C             & 55.3   & 3.1           & 55.2   & 3.1 \\
    $^{124}$Xe                      & 21.0   & 6.3           & 22.6   & 5.0 \\
    $^{83\mathrm{m}}$Kr             & 17.1   & 5.1           & \multicolumn{2}{c}{$17.0^{+3.5}_{-3.3}$} \\
    $^{125}$I                       & 8.9    & 2.7           & 9.8   & 2.3 \\
    Detector ERs                    & 8.5    & 3.4           & \multicolumn{2}{c}{$9.4^{+1.7}_{-1.6}$} \\
    Accidental coincs.              & 2.7    & 0.6           & 2.6    & 0.6 \\
    $^{127}$Xe + $^{125}$Xe         & 1.5    & 0.3           & 1.2    & 0.3 \\
    Atmospheric $\nu$               & \multicolumn{2}{c}{$(1.1 \pm 0.2) \times 10^{-1}$} & \multicolumn{2}{c}{$(1.1 \pm 0.2) \times 10^{-1}$} \\
    $^{8}$B$+hep\,\nu$              & \multicolumn{2}{c}{$(5.7 \pm 0.6) \times 10^{-2}$} & \multicolumn{2}{c}{$(5.7 \pm 0.6) \times 10^{-2}$} \\
    MSSI                            & \multicolumn{2}{c}{$(4.9 \pm 4.9) \times 10^{-3}$} & \multicolumn{2}{c}{$(4.6^{+4.8}_{-4.6}) \times 10^{-3}$}  \\
    Detector NRs                    & \multicolumn{2}{c}{--} & \multicolumn{2}{c}{$[0, 0.118]$} \\
    $\mathcal{L}_{10}^s$ (1000 GeV/c$^2$)              & \multicolumn{2}{c}{--} & \multicolumn{2}{c}{$1.0^{+1.4}_{-0.7}$} \\
    \hline
    Total counts                    & \multicolumn{2}{c}{--} & 1713 & 39 \\
    \hline
    Observed counts                 & \multicolumn{2}{c}{--} & \multicolumn{2}{c}{1710}$\,\,\,\,\,\,\,\,\,\,\,\,$ \\
    \hline \hline
    \end{tabular}
    \label{tab:backgrounds} 
\end{table}

Following the recommendations in Refs.~\cite{Cowan_2011:Asymptotic,DM_parameters:BAXTER2021_Conventions}, the analysis uses a Profile Likelihood Ratio (PLR), using a two-sided test statistic to assess the compatibility between the data and a variety of signal hypotheses, where we scan over mass or mass splitting for a particular Lagrangian or operator.
The full model was  validated using Goodness of Fit (GoF) tests performed in \{S$1c$,~log$_{10}$(S2$c$)\} space and in one-dimensional projections of each axis. 
The tests were performed for each of the three data samples, with none returning a p-value below 0.05.
More details on the fit and expectations for the \emph{prompt veto} and \emph{delayed veto} samples are given in the \textcolor{black}{Supplemental Material}.

\paragraph*{Results -\label{sec:results}}

\begin{figure}
    \centering
    \includegraphics[width=0.95\linewidth]{\plotfolder 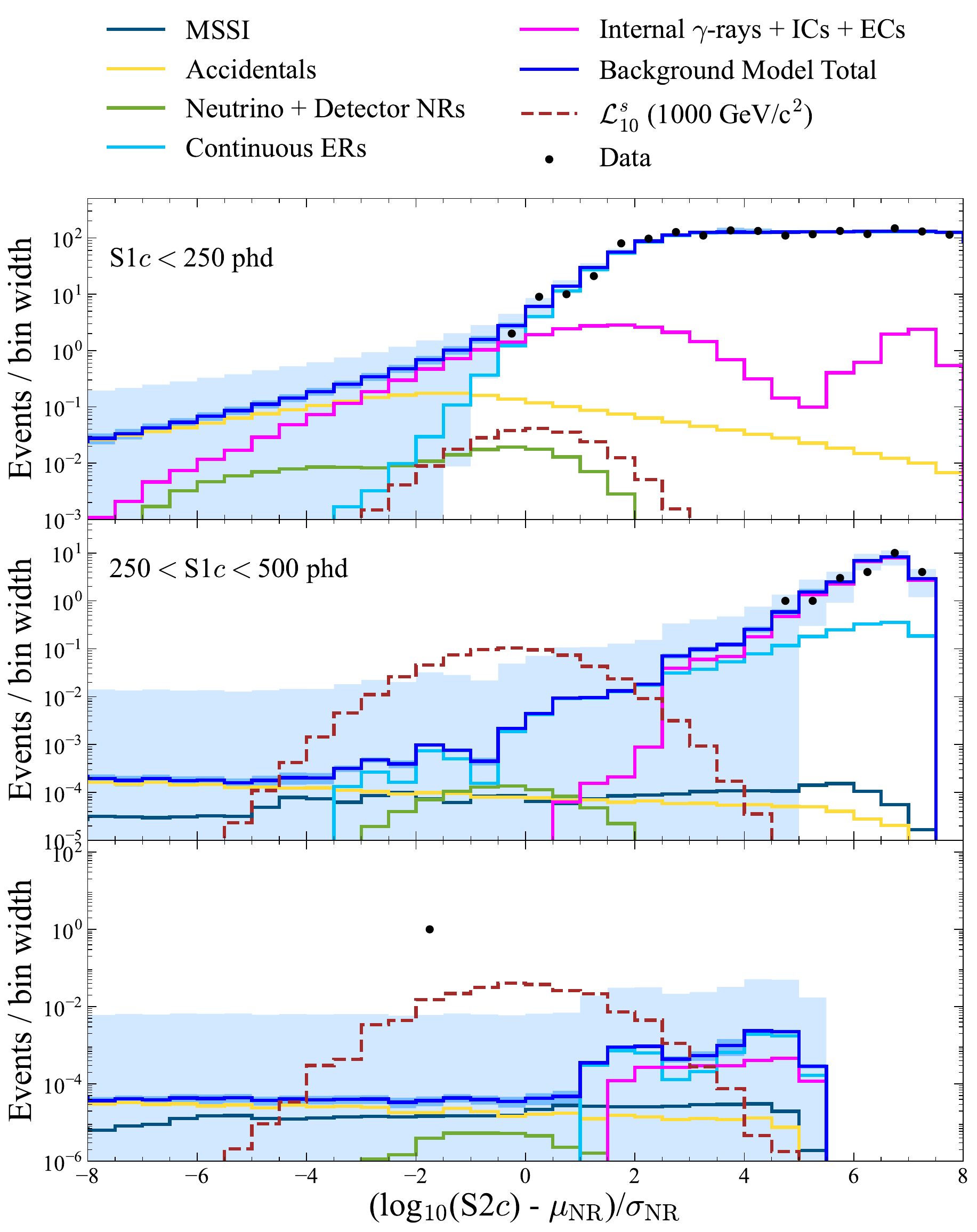}
    \caption{
    Comparison between the data in the \emph{science} sample and the background model corresponding to the post-fit values shown in \autoref{tab:backgrounds} in three S1$c$ ranges.
    Also shown in the dashed brown is the best fit 1000~GeV/c$^2$ $\mathcal{L}^{s}_{10}$ WIMP interaction. 
    The comparison is a projection of the deviation in S2$c$ from the NR median at fixed S1$c$, normalized by the NR band width, and binned in 0.5$\sigma$ bins. The data are plotted at the center of each bin.  
   The total integrated background in the bottom panel is $0.0106\pm0.0008\,(\mathrm{sys})$ counts.
   The shaded dark and light blue bands depict the systematic and statistical uncertainties, respectively, with the latter calculated as the standard deviation of a Poisson distribution with a mean of the central value in the bin. Note that for very low Poisson mean $\mu$, the coverage of the standard deviation is much greater than 68\%, approaching 1-$\mu$ for $\mu << 1$. 
}
    \label{fig:er_projection}
\end{figure}

Driven by the event of interest noted earlier, the p-value to reject the background-only hypothesis is smaller than 1.3$\times10^{-3}$ for a substantial subset of the tested models, corresponding to a local significance greater than 3$\sigma$. 
\autoref{tab:backgrounds} shows the post-fit values of the nuisance parameters for the $\mathcal{L}^{s}_{10}$ model, a representative high-significance model.
Figure~\ref{fig:er_projection} shows the \emph{science} sample data and best-fit model in a 1D projection along the median of the NR band, illustrating the increased ER-NR separation with energy and the low background rate near the event of interest. The local significance for all tested models is shown in Tables~\ref{tab:lagrangian_significance}~and~\ref{tab:operator_significance} in the \textcolor{black}{Supplemental Material}.  

As we are evaluating the sensitivity to a large number of models, a look elsewhere effect (LEE) correction~\cite{DM_parameters:BAXTER2021_Conventions} is calculated using datasets produced by toy Monte Carlo simulations.
The global p-value after the LEE correction is 2.6$\sigma$.
More detail on the LEE implementation is given in the \textcolor{black}{Supplemental Material}.

\begin{figure}
    \centering
    \includegraphics[width=0.9\columnwidth]{\plotfolder 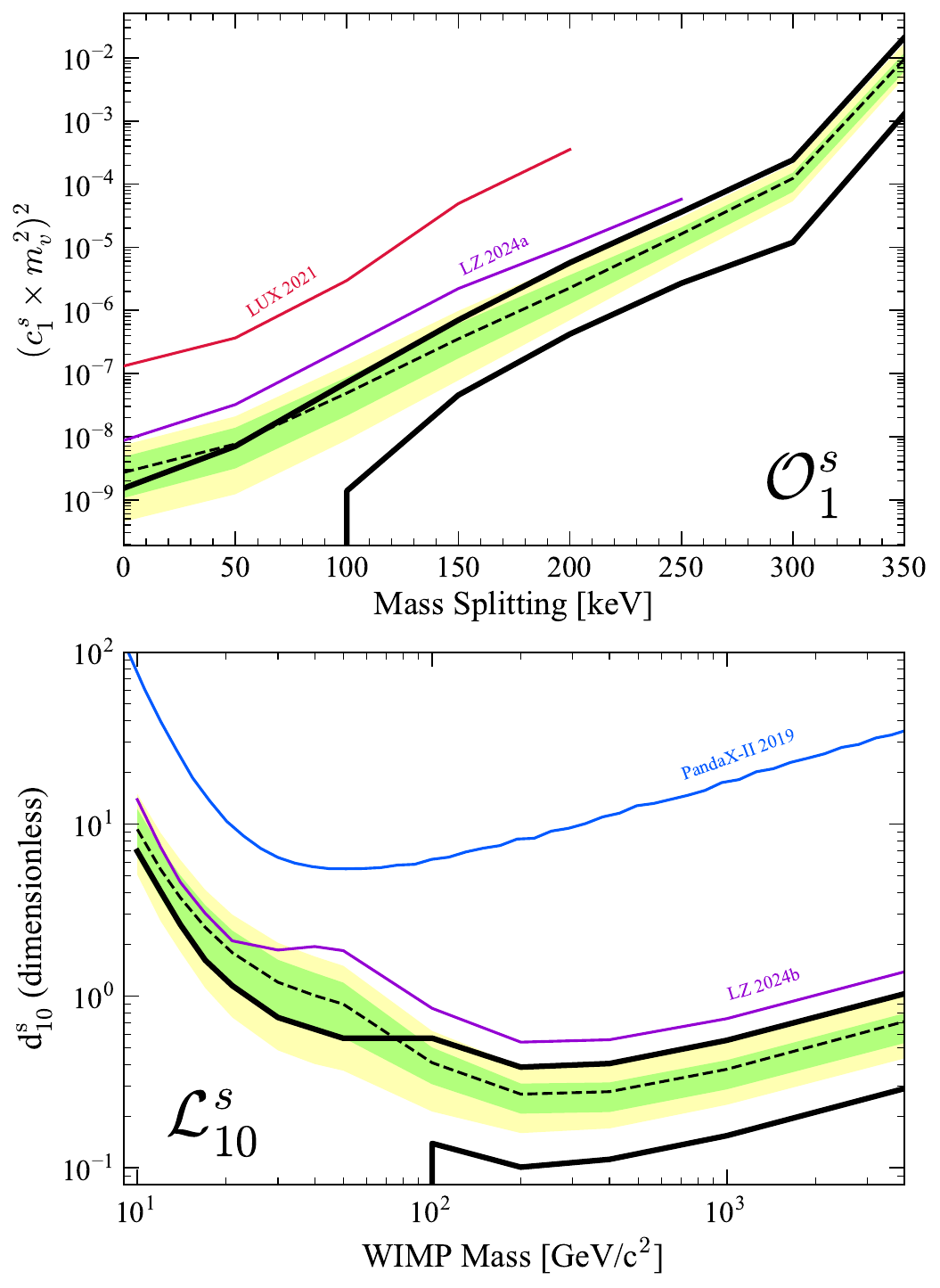}
    \caption{Two-sided 90\% confidence-level intervals (solid-black) of the dimensionless isoscalar WIMP-nucleon coupling.
    \textbf{Top:} Inelastic $\mathcal{O}^{s}_1$ for a 1000~GeV/c$^2$ dark matter mass, scanning over mass splitting.
    \textbf{Bottom:} Elastic $\mathcal{L}^{s}_{10}$, scanning over dark matter mass.
    The lower limit lifts off from zero because of the presence of the event of interest. The black dashed line is the median sensitivity, and the green and yellow bands contain 68\% and 95\% of expected upper limits given the best-fit background model.
    Previous limits are shown in violet from LZ~\cite{LZ:SR1_NREFT_2023,LZ:SR1_NREFT_Lagrangian_2024}, red from LUX~\cite{LUX:EFTR4_2021}, and blue from PandaX~\cite{PandaX2:SD_EFT_2019}.
    }
    \label{fig:limit_plot}
\end{figure}

We construct 90\% confidence level intervals on all tested models as a function of WIMP mass, with the two example models shown in \autoref{fig:limit_plot}. 
For $\mathcal{L}_{1}-\mathcal{L}_{20}$, the quantity constrained is the interaction coupling parameter, $d_j$, as a function of WIMP mass.
For $\mathcal{O}_{1}$ and $\mathcal{O}_{4}$ the coupling strength limits are cast in the dimensionless form $(c^N_i\times m^2_\nu)^2$, where $m_\nu$ is the Higgs vacuum expectation value~\cite{Anand:MathematicaEFT}. 
Given the observed local significance, the lower limit is non-zero for some masses.
The derived upper limits set world-leading constraints on all models tested. 
Confidence intervals for all models can be found in the \textcolor{black}{Data Release}.

\paragraph{Discussion --\label{sec:discussion}}
Given the 3$\sigma$-level local significance, increased scrutiny of the analysis is warranted. We discuss here detector conditions around the event of interest and additional details of key background topologies.
Information regarding analysis of the event waveform is given in the \textcolor{black}{Supplemental Material}.

The event occurred in close proximity to a calibration period. On 16 June 2023, an ER calibration source, $^{57}$Co, was deployed in a calibration source deployment (CSD) tube~\cite{LZ:Calibrations_2024} and removed 25 minutes before the event was recorded. The source deployment was on the opposite side of the detector from the event of interest. Used for calibrations of the Skin, $^{57}$Co emits primarily 122~keV and 136~keV $\gamma$-rays with a mean free path of less than 4 mm in LXe.
The previous NR calibration, one of three in the full dataset, was an AmBe deployment on 8 June 2023. 
The calibration data and the science data in the periods around the event of interest were analyzed and no anomalous populations were observed. The previous observed muon in LZ occurred in the OD 41 minutes before the event, with the previous muon in the TPC occurring 127 minutes before.
Other key detector parameters showed no abnormalities in the scintillation and ionization responses of the detector in the temporal and spatial vicinity of the interaction in question. LZ has demonstrated the ability to tag events as being likely $^{214}$Pb decays via a radon tag~\cite{LZ:2025xxf}, but the tag cannot be applied to the event of interest as it occurred while the detector was in the mixed flow state~\cite{LZ:SR3_WS2024}.

Although most events in the WS~ROI are from ER backgrounds, discrimination against ER events improves at higher energies with very little leakage predicted for S1$c>250$~phd, as illustrated in~\autoref{fig:er_projection}. Electron capture decays have increased recombination, and hence more S1 and less S2 relative to $\beta$ decays~\cite{LZ:2025hud}.  $^{124}$Xe and $^{125}$I have double-vacancy decay modes releasing 64.3~keV and 67.3~keV of energy, respectively, similar to the reconstructed energy of the event of interest if interpreted as a pure ER. These decays are modeled following Ref.~\cite{LZ:2025hud}, and there is a systematic uncertainty in projecting into the tail of the distribution that we do not include in the statistical inference. We note the lack of events between the event of interest and the bottom of the ER band in~\autoref{fig:finaldataset-S1-logS2}.

In the region of \{S1$c$, log$_{10}$(S2$c$)\} space where the event occurred, the dominant backgrounds are accidentals, atmospheric neutrinos, and MSSI events. 
While these processes are able to produce a high energy NR-like interaction, such an event is generally in the tail of a larger distribution. 
The potential for significant mismodeling of these backgrounds is thus constrained by the lack of a larger population in each case.

Accidental coincidences are expected to be most prevalent at low energies,
and the accidentals model is well tested in previous analyses~\cite{LZ:SR3_WS2024, LZ:2025lowmass}. More detail can be found in the \textcolor{black}{Supplemental Material}.
While the flux of atmospheric neutrinos carries some uncertainty, the coherent scattering neutrino interaction is well understood, producing an approximately exponentially-decaying recoil spectrum with increasing energy.
In addition, the reconstructed energy of the event is not consistent with scattering coherence.

The MSSI simulation is validated using a larger 5.4 tonne analysis volume, a higher energy sideband (HE~SB: 800 $<$ S1$c$ $<$ 1700~phd, $10^{2.75}$ $<$ S2$c$ $<$ $10^{4.3}$~phd), and the \emph{prompt veto} sideband. The 5.4 tonne analysis volume uses the same $z$ boundaries as Ref.~\cite{LZ:SR3_WS2024} with a radial boundary defined by requiring an expected wall background count of less than $(1.0 \pm 0.5)\times10^{-2}$ in the $3-600$~phd WS~ROI; the radial boundary in Ref.~\cite{LZ:SR3_WS2024} was defined by the same criterion over the $3-80$~phd ROI used in that analysis. 
Events are categorized as wall- or RFR-like, \emph{science} or \emph{prompt veto}, WS~ROI or HE~SB, and in one of the two analysis volumes; excluding the search space of the WS~ROI in the 4.7 tonne FV leaves 12 bins of comparison. 
A binned Poisson likelihood chi-square test across those 12 bins returns a p-value of 0.7, indicating no inconsistency between model and data~\cite{BAKER1984437}.
More detail on the MSSI model validation is provided in the \textcolor{black}{Supplemental Material}.

Neutrons are generally considered the most likely cause of NR interactions, and we explore three primary sources: ($\alpha$,n), spontaneous fission, and muon-induced cascades. 
A neutron must have at least 8 MeV in order to impart 250~keV to a xenon nucleus in elastic scattering, and the cross section for elastic scattering of neutrons on xenon becomes highly forward peaked as the neutron energy increases, leading to small energy depositions. 
Similar to neutrinos and accidental coincidences, it is thus very unlikely to observe a single  recoil at this energy without also observing several more events at lower energies.
This spectral information, in addition to the veto systems, provides a strong constraint in the statistical inference on the contribution of neutrons to the background in the region of the event of interest. 
More details regarding 
spontaneous fission and muon-induced 
neutrons are presented in the \textcolor{black}{Supplemental Material}.

\paragraph{Conclusion --\label{sec:conclusion}}

This Letter reports results from the LZ experiment exploring an extended nuclear recoil energy region of 5.4 to 270~keV, motivated by expected recoil energy spectra for WIMP--nucleon interactions in NREFT~\cite{Fitzpatrick:EFT}.
After event and fiducial volume selection in a data sample corresponding to a 2.84 tonne-year exposure, we observe one event consistent with an elastic nuclear recoil interaction with a recoil energy of $248\pm23\,\mathrm{(stat)}\pm23\,\mathrm{(sys)}$~keV. 
The event can be described by signal spectra predicted by several NREFT Lagrangians or inelastic DM models, resulting in a global significance of 2.6$\sigma$ and a maximum local significance of 3.4$\sigma$, typically for WIMP masses above 200~GeV/c$^2$.
Several rare background processes and detector effects were studied in detail but none could be identified as a likely explanation for the event of interest. LZ has continued to take data with the same electric field conditions since 1 April 2024.

\paragraph{\label{sec:acknowledgements}Acknowledgements --}

The research supporting this work took place in part at the Sanford Underground Research Facility (SURF) in Lead, South Dakota. Funding for this work is supported by the U.S. Department of Energy, Office of Science, Office of High Energy Physics under Contract Numbers DE-AC02-05CH11231, DE-AC02-76SF00515, DE-AC52-07NA27344, DE-SC0008475, DE-SC0010004, DE-SC0010010, DE-SC0010072, DE-SC0011702, DE-SC0012447, DE-SC0012704, DE-SC0014223, DE-SC0015535, DE-SC0015910, DE-SC0018982, DE-SC0019066, DE-SC0019193, DE-SC0020216, DE-SC0025446, DE-SC0025629, DE-SC0026014, DE-SC0026544. This research was also supported by U.S. National Science Foundation (NSF) ; the UKRI’s Science \& Technology Facilities Council under award numbers ST/M003639/1, ST/R003181/1, ST/W000466/1, ST/W000490/1, ST/W000555/1, ST/W00058X/1, ST/W000628/1, ST/W000636/1, ST/X00600X/1, ST/X006050/1, UKRI2837, UKRI2840, UKRI2841, UKRI2842, UKRI2843, UKRI2844, UKRI2847, UKRI2851, UKRI2852; the Portuguese Foundation for Science and Technology (FCT) under award number PTDC/FIS-PAR/2831/2020; the Institute for Basic Science, Korea (budget number IBS-R016-D1); the Swiss National Science Foundation (SNSF) under award number 10001549; the European Union’s research and innovation programme Horizon Europe under the Marie Skłodowska-Curie grant agreement No 101209429. This research was supported by the Australian Government through the Australian Research Council Centre of Excellence for Dark Matter Particle Physics under award number CE200100008. We acknowledge additional support from the UK Science \& Technology Facilities Council (STFC) for PhD studentships and the STFC Boulby Underground Laboratory in the U.K., the GridPP~\cite{faulkner2005gridpp,britton2009gridpp} and IRIS Collaborations, in particular at Imperial College London and additional support by the University College London (UCL) Cosmoparticle Initiative, and the University of Zurich. We acknowledge additional support from the Center for the Fundamental Physics of the Universe, Brown University. K.T. Lesko acknowledges the support of the Royal Society, the Wolfson Foundation, and Brasenose College. This research used resources of the National Energy Research Scientific Computing Center, a DOE Office of Science User Facility supported by the Office of Science of the U.S. Department of Energy under Contract No. DE-AC02-05CH11231. We gratefully acknowledge support from GitLab through its GitLab for Education Program. Fermilab is managed by Fermi Forward Discovery Group, LLC, acting under Contract No. 89243024CSC000002. The University of Edinburgh is a charitable body, registered in Scotland, with the registration number SC005336. The assistance of SURF and its personnel in providing physical access and general logistical and technical support is acknowledged. We acknowledge the South Dakota Governor's office, the South Dakota Community Foundation, the South Dakota State University Foundation, and the University of South Dakota Foundation for use of xenon. We also acknowledge the University of Alabama for providing xenon. For the purpose of open access, the authors have applied a Creative Commons Attribution (CC BY) license to any Author Accepted Manuscript version arising from this submission. Finally, we respectfully acknowledge that we are on the traditional land of Indigenous American peoples and honor their rich cultural heritage and enduring contributions. Their deep connection to this land and their resilience and wisdom continue to inspire and enrich our community. We commit to learning from and supporting their effort as original stewards of this land and to preserve their cultures and rights for a more inclusive and sustainable future.

\bibliography{references}

\clearpage
\onecolumngrid
\appendix
\section*{Supplemental Material}
\pagestyle{plain}
\setcounter{figure}{0}
\renewcommand{\thefigure}{S\arabic{figure}}

\setcounter{table}{0}
\renewcommand{\thetable}{S\arabic{table}}

\subsection{Data Samples and Salt}
\begin{figure}[!h]
    \includegraphics[width=0.41\columnwidth]{\plotfolder 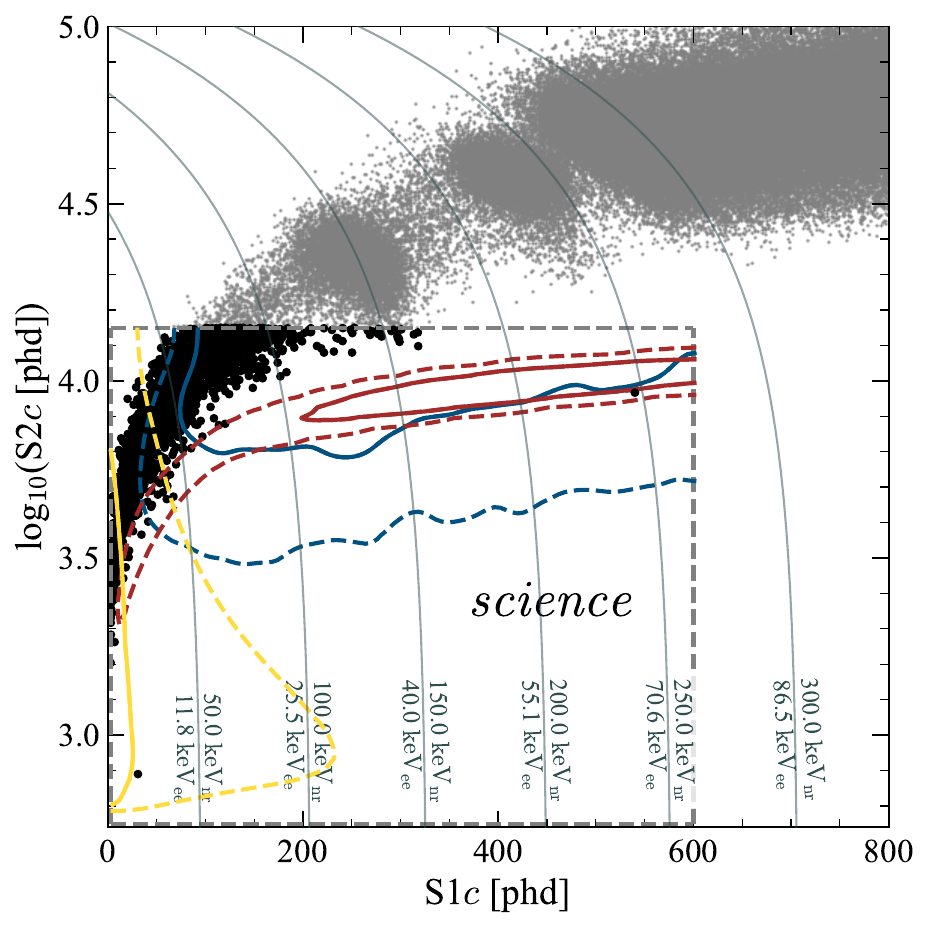}
    \includegraphics[width=0.41\columnwidth]{\plotfolder 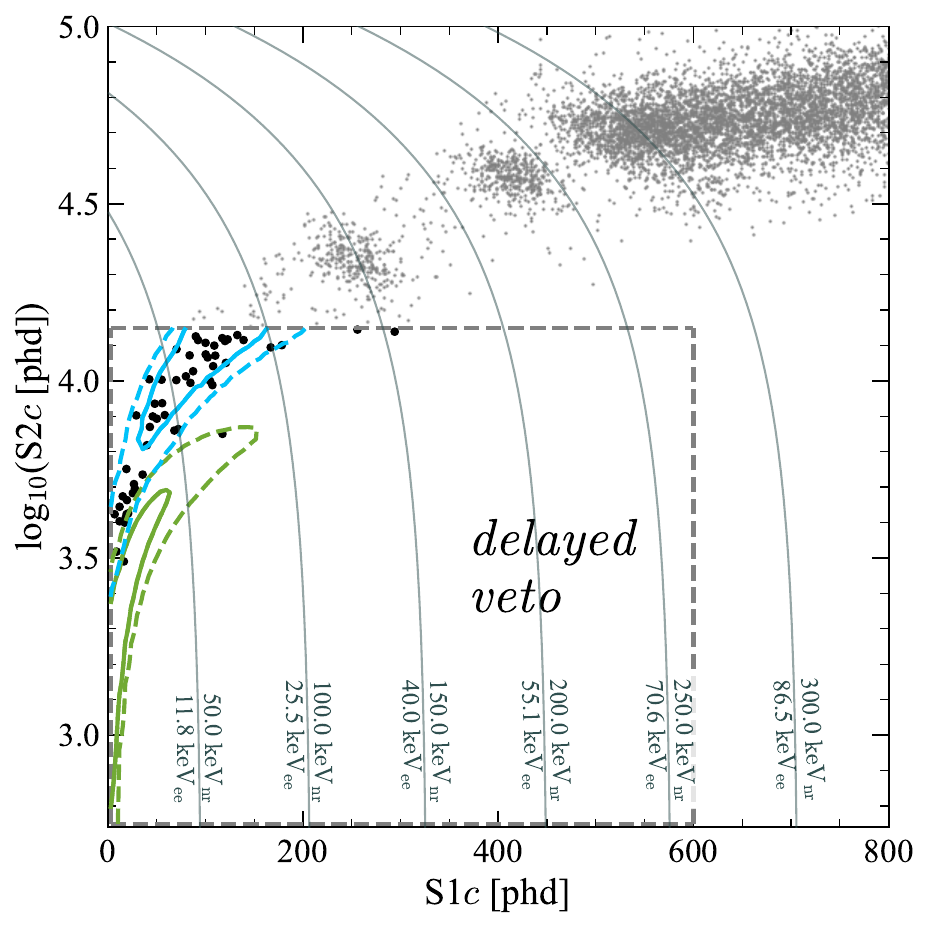}
    \includegraphics[width=0.41\columnwidth]{\plotfolder 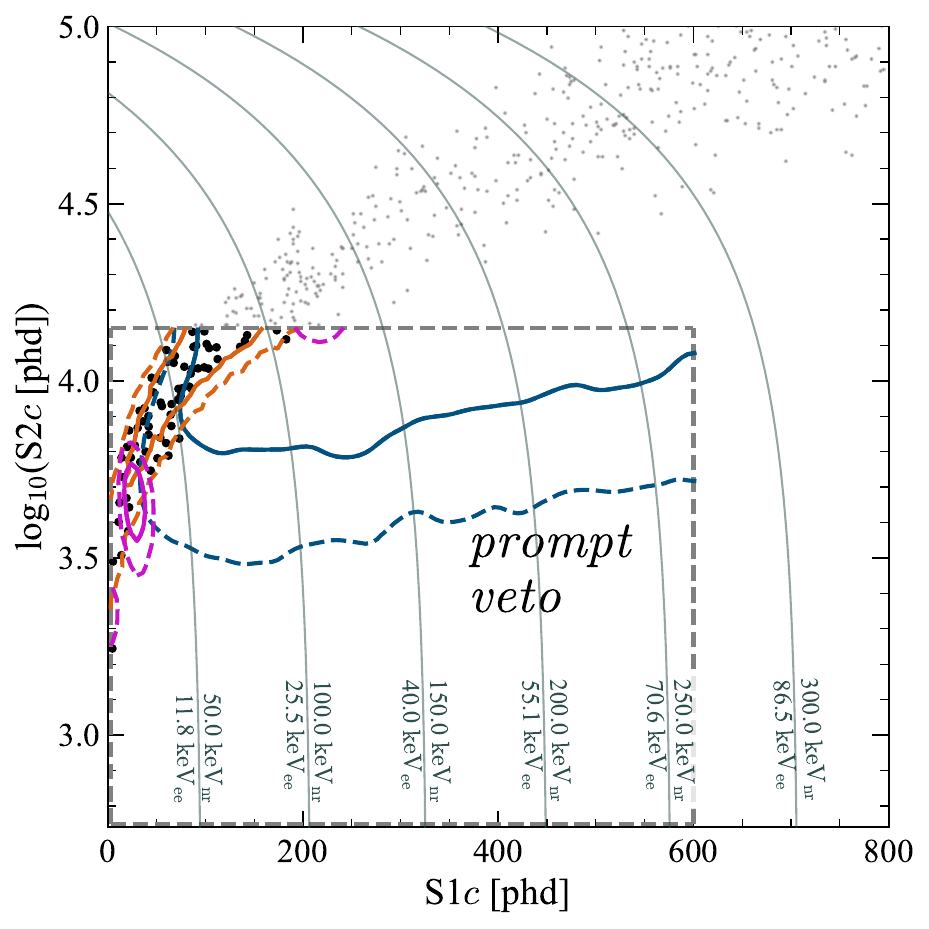}
    \includegraphics[width=0.41\columnwidth]{\plotfolder 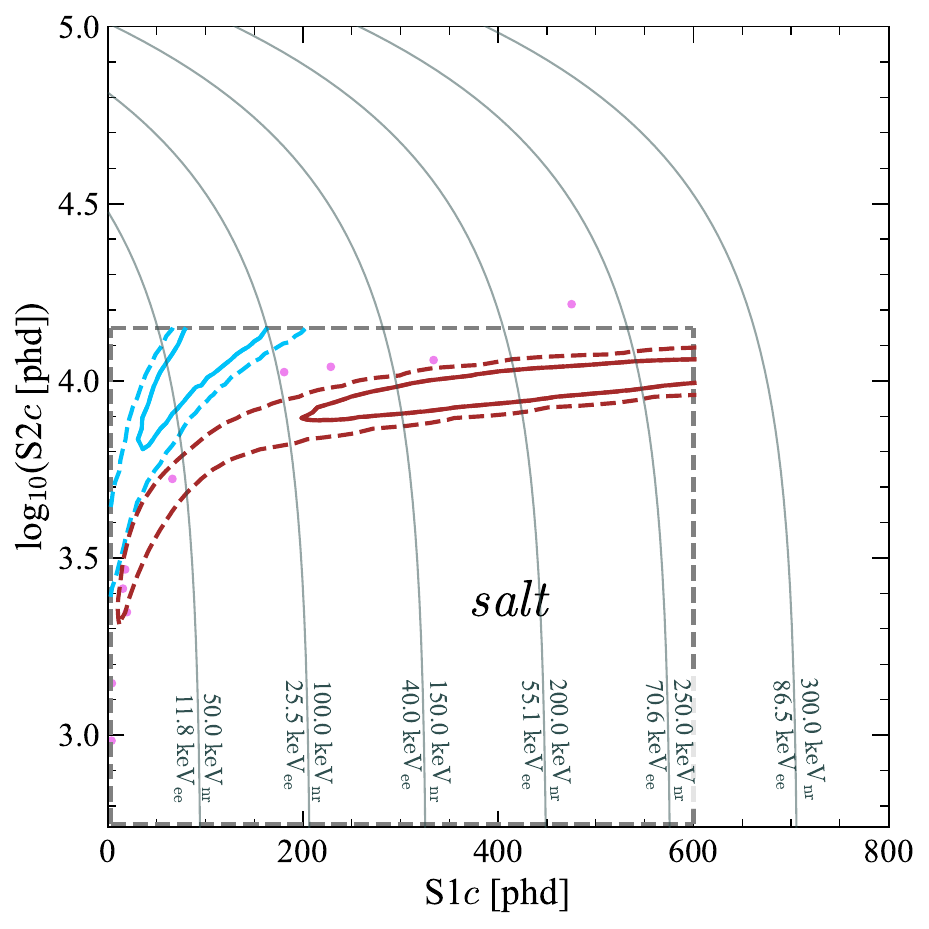}
    \caption{
    The black points show the WIMP search data after all cuts in \{S1c, log$_{10}$(S2c)\} space with each panel showing a different sample, including the \emph{salt} (pink) in the bottom right.
    The dashed gray area indicates the WS~ROI.
    The contours show 68\% (solid) and 95\% (dashed) containment regions for: a 1000~GeV/c$^2$ $\mathcal{L}^{s}_{10}$ interaction (brown), MSSI (dark blue), accidental coincidences (yellow), combined ER backgrounds (light blue), and the detector NR background (green). The lower left panel splits the ER contribution into ERs from detector materials (orange) and $^{127}$Xe (magenta).  
    Contours of constant recoil energy are shown in gray.
    The \emph{science} sample is the same data as those shown in~\autoref{fig:finaldataset-S1-logS2}.
    }
    \label{fig:all_samples_data}
\end{figure}

\begin{figure}
    \centering
    \includegraphics[width=0.5\linewidth]{\plotfolder 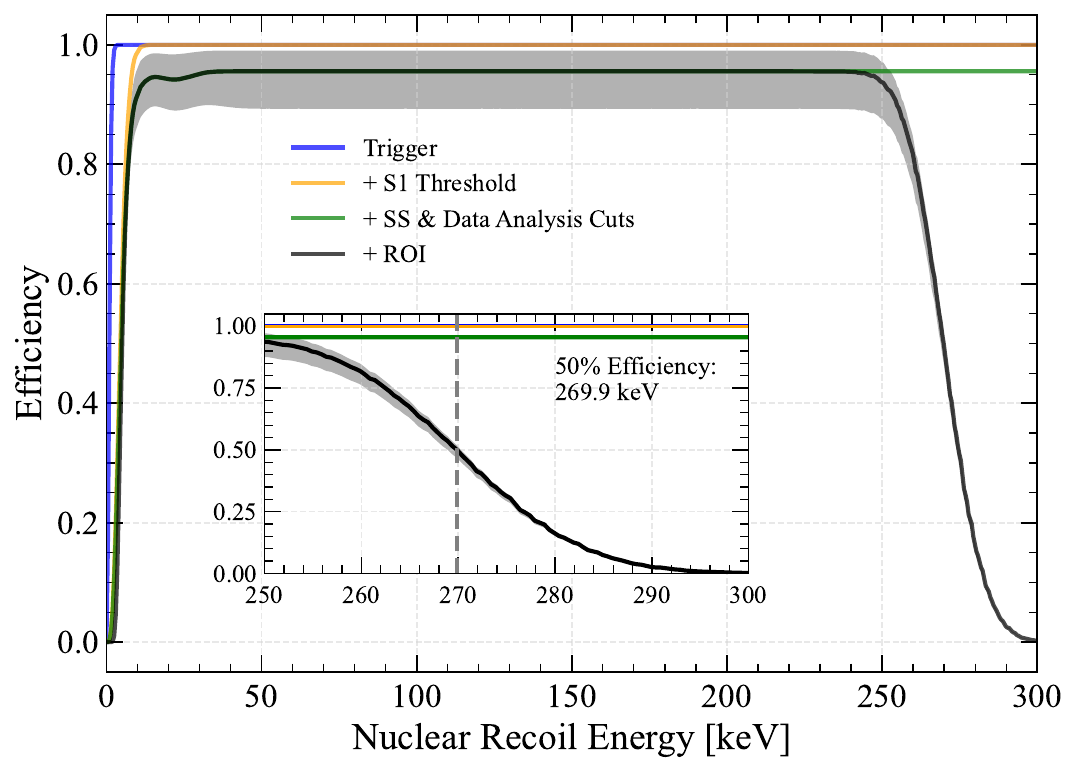}
    \caption{
    Signal efficiency as a function of the NR energy from the trigger (blue); S1 threshold (orange); the SS reconstruction and analysis cuts (green); and the WS~ROI in S1$c$ and S2$c$ (black).
    The high energy behavior is shown in the insert, where the dotted line at 269.9~keV indicates the nuclear recoil energy at which the efficiency is 50\%.
    The uncertainty (gray) is assessed with $^3$H, $^{220}$Rn, and AmLi calibration data. The low energy behavior is shown in Fig.~2 of Ref.~\cite{LZ:SR3_WS2024}, crossing 50\% efficiency at 5.4~keV.
    }
    \label{fig:efficiency}
\end{figure}

Figure~\ref{fig:all_samples_data} shows the events passing all data selection criteria separated into the three samples used in the statistical inference, plus a fourth sample showing the salt. 
The points in the \emph{delayed veto} sample come primarily from random coincidences of background pulses in the veto detectors with ER backgrounds in the TPC due to the long coincidence window. 
The points in the \emph{prompt veto} sample come primarily from $\gamma$-rays interacting in the top corner of the TPC and in the Skin detector (see Fig.~\ref{fig:roi_fv}).
The bottom right panel shows the \emph{salt} events, which appeared in the \emph{science} sample prior to unsalting. 
In addition to the seven salt events from Ref.~\cite{LZ:SR3_WS2024}, three new salt events appear in the WS~ROI, with a fourth just above. 
As discussed in the main text, the model used to generate the salt was tuned before the AmBe calibration data set was obtained, and the salt events fall above the main signal region in brown. Figure~\ref{fig:efficiency} shows the signal detection efficiency as a function of NR energy.

\subsection{Log-Likelihood Function and the Look Elsewhere Effect}
The likelihood is defined as
\begin{equation}
\begin{aligned}
    \mathcal{L}(\theta) = \prod_{i=0}^{2} \Bigg[ \mathrm{Pois}(N_i|\mu_{\mathrm{tot},i})
     \times \prod^{N_i}_{e=1}\frac{1}{\mu_{\mathrm{tot},i}}\bigg(\mu_{\mathrm{sig},i}(\vec{\lambda})f_{\mathrm{sig},i}(\vec{x}_e) + \sum^{N_b}_{b=1}\mu_{b,i}(\vec{\lambda}) f_{b,i}(\vec{x}_e) \bigg) \Bigg] 
     \times \prod^{N_b+N_\lambda}_{k=0} g_k(\phi_k|\vec{\nu}_{\phi_k}),
\end{aligned}
\label{eq:likelihood_ack}
\end{equation}
where $N_i$ is the number of data events in the $i$th sample; $\mu_{\rm{tot},i}$ is the expected number of events expressed as the sum of the expected number of signal ($\mu_{\mathrm{sig},i}$) and background ($\mu_{b,i}$) events and both are a function of $\theta$, which expresses the parameters of the model. The numbers of signal and background events in each sample are not independent between samples, but are related with tagging efficiency ($\vec{\lambda}$) parameters. The vector $\vec{x}_e$ is the event coordinate in \{S1c, log$_{10}$(S2c)\} space.
The functions $f_{\mathrm{sig},i}$ and $f_{b,i}$ describe the probability density functions of the signal and background components of the model. 
The $g(\phi_k|\vec{\nu}_{\phi_k})$ term represents constraint functions on the rate of each background component and the tagging efficiencies $\vec{\lambda}$, where $\vec{\nu_\phi}$ defines any fixed parameters of the function. 

Each background is constrained by a Gaussian function except for the detector neutrons, which are not given a direct rate constraint.
The neutron component is constrained by both the \emph{prompt veto} and \emph{delayed veto} by tagging efficiencies $\lambda_\text{PN}$ and $\lambda_\text{DN}$, respectively.
The \emph{prompt veto} sample additionally constrains the MSSI component through a tagging efficiency, denoted $\lambda_\text{MSSI}$, and single scatters from Detector ERs and $^{127}$Xe/$^{125}$Xe through $\lambda_\text{PG}$.

Goodness-of-fit (GoF) tests were performed to study the compatibility of the background model with the data, including one-dimensional projections on S1$c$ and log$_{10}$(S2$c$), and over a two-dimensional binning into bins containing equal expectation values of 21.1, 5.6, and 7.7 for the \emph{science, prompt veto,} and \emph{delayed veto} samples, respectively.
All tests returned p-values above 0.05, and were thus considered acceptable. Pull-studies comparing the best-fit parameters with the expected spread found that all parameters were pulled by less than $2\sigma$ from their nominal values, and no parameters were identified as being strongly correlated with the signal strength. 
The GoF tests support the validity of the background model in the WS~ROI, but because of their binned nature, they are not sensitive to single outlier events such as the event of interest. 
Outlier-focused rejection tests if applied would also not have rejected this event since it lies in the signal region where an analysis should not reject valid events.

A look elsewhere effect (LEE) correction is calculated for this analysis using toy Monte Carlo datasets, as described in Ref.~\cite{DM_parameters:BAXTER2021_Conventions}.
Toy datasets are generated from the background-only model and analyzed with the same procedure as the real data. 
For each toy dataset, the smallest p-value from all of the models is recorded, and a probability distribution of minimum p-values is generated. 
The minimum p-value for the real data is compared to this distribution to extract a global p-value that quantifies how likely it is to observe a data set at least as extreme as the real data, given the model space tested.

The total LEE depends on the signal search space, and care was taken to span the full range of signal search space while also not performing redundant calculations on very similar spectra.
The space of models is chosen to cover all of the isoscalar and isovector options $\mathcal{L}_1$--$\mathcal{L}_{20}$ using 13 logarithmically spaced dark matter masses from 10 to 4000 GeV/c$^2$. 
The inelastic models include isoscalar and isovector options for $\mathcal{O}_1$ and $\mathcal{O}_4$ for 400~GeV/c$^2$, 1000~GeV/c$^2$, and 4000~GeV/c$^2$ WIMP masses and eight mass splittings: 0, 50, 100, 150, 200, 250, 300, 350~keV. 
In total, there are 616 models. 
However, the model space has some degeneracies, and as the LEE only depends on the number of distinguishable spectra, these degeneracies have been exploited to reduce the computation time.
Specifically, the pairs $\mathcal{L}_1$ and $\mathcal{L}_5$; $\mathcal{L}_2$ and $\mathcal{L}_8$; $\mathcal{L}_3$ and $\mathcal{L}_{17}$; $\mathcal{L}_4$ and $\mathcal{L}_{20}$; and $\mathcal{L}_{11}$ and $\mathcal{L}_{14}$ differ from each other only by a scalar constant. 
In addition, for some models, the energy spectra for the isoscalar and isovector couplings of that model are indistinguishable; this applies to Lagrangians $\mathcal{L}_2$, $\mathcal{L}_4$, $\mathcal{L}_7$, $\mathcal{L}_8$, $\mathcal{L}_{10}$, $\mathcal{L}_{11}$, $\mathcal{L}_{14}$, $\mathcal{L}_{15}$, $\mathcal{L}_{19}$, and $\mathcal{L}_{20}$ and the inelastic operator $\mathcal{O}_4$.
Finally, for all of the Lagrangians, masses of  400~GeV/c$^2$ and above have nearly degenerate recoil spectra, so only one mass (1000~GeV/c$^2$) is kept. 
This procedure reduces the number of models to be tested to 293.

When reporting our results, we perform a raster scan over masses or masses and mass splittings as appropriate for the Lagrangian or operator being tested. Our limits therefore show the best region of coupling strength for each mass separately, and not a region optimized in coupling and mass simultaneously. We do not set an additional threshold requirement on the local significance of any model to report a lower limit on the coupling strength in this scan.

\subsection{Fit Results in Veto Samples}
In this section we present the pre- and post-fit expectations in the \emph{prompt} (\autoref{tab:prompt_sample_fit_table}) and \emph{delayed} (\autoref{tab:delayed_sample_fit_table}) samples for the best combined fit of the background model plus a 1000~GeV/c$^2$ $\mathcal{L}_{10}^s$ signal to the data. 

\setlength{\tabcolsep}{12pt} 
\begin{table}[ht]
    \centering
    \caption{\textbf{\emph{Prompt} sample:}
    The expected and fitted numbers of events from sources considered in the 220~d~$\times$~4.71~t exposure.
    The middle column contains the predicted number of events with uncertainties as described in the text.
    These uncertainties are used as constraints in a combined fit of the background model and a 1000~GeV/c$^2$ $\mathcal{L}^{s}_{10}$ WIMP signal to the data.
    The result of the fit is shown in the right column.}
    \begin{tabular}{l
            r@{$\,\pm\,$}l
            r@{$\,\pm\,$}l}
     \hline 
     \hline 
     Source & \multicolumn{2}{c}{Expected Events} & \multicolumn{2}{c}{Fit Result} \\ \hline
     Internal $\beta$ decays          & \multicolumn{2}{c}{$(1.3 \pm 0.2) \times 10^{-1}$} & \multicolumn{2}{c}{$(1.33 \pm 0.04) \times 10^{-1}$} \\
     $\nu$ ER                         & \multicolumn{2}{c}{$(1.41 \pm 0.08) \times 10^{-2}$} & \multicolumn{2}{c}{$(1.40 \pm 0.08) \times 10^{-2}$} \\
     $^{136}$Xe                       & \multicolumn{2}{c}{$(1.1 \pm 0.2) \times 10^{-2}$}  & \multicolumn{2}{c}{$(1.1 \pm 0.2) \times 10^{-2}$} \\
     CH$_3$T \& $^{14}$C              & \multicolumn{2}{c}{$(5.5 \pm 0.3) \times 10^{-3}$}     & \multicolumn{2}{c}{$(5.5 \pm 0.3) \times 10^{-3}$} \\
     $^{124}$Xe                      & \multicolumn{2}{c}{$(2.1 \pm 0.6) \times 10^{-3}$} & \multicolumn{2}{c}{$(2.3 \pm 0.5) \times 10^{-3}$} \\
     $^{83\mathrm{m}}$Kr             & \multicolumn{2}{c}{$(1.7 \pm 0.5) \times 10^{-3}$} & \multicolumn{2}{c}{$\left(2.0^{+0.4}_{-0.3}\right) \times 10^{-3}$} \\
     $^{125}$I                       & \multicolumn{2}{c}{$(8.9 \pm 2.7) \times 10^{-4}$} & \multicolumn{2}{c}{$(9.8 \pm 2.3) \times 10^{-4}$} \\
     Detector ERs                    & $\,\,\,$62.2     & 24.9   & \multicolumn{2}{c}{$61.6^{+11.5}_{-10.5}$}$\,\,\,\,\,\,$ \\
     Accidental coincidences         & \multicolumn{2}{c}{$(2.7 \pm 0.6) \times 10^{-4}$} & \multicolumn{2}{c}{$(2.6 \pm 0.6) \times 10^{-4}$} \\
     $^{127}$Xe + $^{125}$Xe         & 11.2     & 2.4    & $\,\,\,\,\,\,$7.7 & 2.1 \\
     Atmospheric $\nu$               & \multicolumn{2}{c}{$(1.1 \pm 0.2) \times 10^{-5}$} & \multicolumn{2}{c}{$(1.1 \pm 0.2) \times 10^{-5}$} \\
     $^{8}$B$+hep\,\nu$              & \multicolumn{2}{c}{$(5.7 \pm 0.6) \times 10^{-6}$} & \multicolumn{2}{c}{$(5.7 \pm 0.6) \times 10^{-6}$} \\
     MSSI                            & \multicolumn{2}{c}{$(7.7 \pm 7.7) \times 10^{-2}$} & \multicolumn{2}{c}{$(7.1^{+7.6}_{-7.1}) \times 10^{-2}$} \\
     Detector NRs                    & \multicolumn{2}{c}{--} & \multicolumn{2}{c}{$[0, 0.07]$}  $\,\,\,\,\,\,\,\,\,\,$\\
     $\mathcal{L}_{10}^s$ (1000 GeV/c$^2$)                & \multicolumn{2}{c}{--} & \multicolumn{2}{c}{$\left(1.0^{+1.4}_{-0.7}\right) \times 10^{-4}$} \\
     \hline
     $\lambda_{\text{DN}}$           & \multicolumn{2}{c}{$(8.7 \pm 0.2) \times 10^{-1}$} & \multicolumn{2}{c}{$(8.7 \pm 0.1) \times 10^{-1}$} \\
     $\lambda_{\text{PN}}$           & \multicolumn{2}{c}{$(5.0 \pm 1.0) \times 10^{-2}$} & \multicolumn{2}{c}{$(5.0 \pm 0.7) \times 10^{-2}$} \\
     
     $\lambda_{\text{PG}}$           & \multicolumn{2}{c}{$(8.8 \pm 0.2) \times 10^{-1}$} & \multicolumn{2}{c}{$\left(8.7 \pm 0.2\right) \times 10^{-1}$} \\
     $\lambda_{\text{MSSI}}$         & \multicolumn{2}{c}{$(0.94 \pm 0.2) \times 10^{-1}$} & \multicolumn{2}{c}{$(9.4 \pm 0.2) \times 10^{-1}$} \\
     \hline
     \hline
     Total counts                    & \multicolumn{2}{c}{--} & 69.4 & 11.7 \\
     \hline
     Observed counts                 & \multicolumn{2}{c}{--} & \multicolumn{2}{c}{66} $\,\,\,\,\,\,\,\,\,$\\
     \hline \hline
    \end{tabular}
    \label{tab:prompt_sample_fit_table} 
\end{table}

\setlength{\tabcolsep}{12pt} 
\begin{table}[ht]
    \centering
    \caption{\textbf{\emph{Delayed} sample:}
    The expected and fitted numbers of events from sources considered in the 220~d~$\times$~4.71~t exposure.
    The middle column contains the predicted number of events with uncertainties as described in the text.
    These uncertainties are used as constraints in a combined fit of the background model and a 1000~GeV/c$^2$ $\mathcal{L}^{s}_{10}$ WIMP signal to the data.
    The result of the fit is shown in the right column.}
    \begin{tabular}{l
            r@{$\,\pm\,$}l
            r@{$\,\pm\,$}l}
     \hline 
     \hline 
     Source & \multicolumn{2}{c}{Expected Events} & \multicolumn{2}{c}{Fit Result} \\ \hline
     Internal $\beta$ decays          & 39.5    & 4.7            & 39.4    & 1.3 \\
     $\nu$ ER                         & 4.1     & 0.2            & 4.1     & 0.2 \\
     $^{136}$Xe                       & 3.2     & 0.5            & 3.4     & 0.5 \\
     CH$_3$T \& $^{14}$C              & 1.6     & 0.1            & 1.6     & 0.1 \\
     $^{124}$Xe                      & \multicolumn{2}{c}{$(6.2 \pm 1.9) \times 10^{-1}$} & \multicolumn{2}{c}{$(6.6 \pm 1.5) \times 10^{-1}$} \\
     $^{83\mathrm{m}}$Kr             & \multicolumn{2}{c}{$(5.0 \pm 1.5) \times 10^{-1}$} & \multicolumn{2}{c}{$(5.0 \pm 1.0) \times 10^{-1}$} \\
     $^{125}$I                       & \multicolumn{2}{c}{$(2.6 \pm 0.8) \times 10^{-1}$} & \multicolumn{2}{c}{$(2.9 \pm 0.7) \times 10^{-1}$} \\
     Detector ERs                    & \multicolumn{2}{c}{$(2.5 \pm 1.0) \times 10^{-1}$} & \multicolumn{2}{c}{$(2.8 \pm 0.5) \times 10^{-1}$} \\
     Accidental coincidences         & \multicolumn{2}{c}{$(8.0 \pm 1.9) \times 10^{-2}$} & \multicolumn{2}{c}{$(7.7 \pm 1.8) \times 10^{-2}$} \\
     $^{127}$Xe + $^{125}$Xe         & \multicolumn{2}{c}{$(4.5 \pm 1.0) \times 10^{-2}$} & \multicolumn{2}{c}{$(3.4 \pm 0.9) \times 10^{-2}$} \\
     Atmospheric $\nu$               & \multicolumn{2}{c}{$(3.3 \pm 0.7) \times 10^{-3}$} & \multicolumn{2}{c}{$(3.3 \pm 0.7) \times 10^{-3}$} \\
     $^{8}$B$+hep\,\nu$              & \multicolumn{2}{c}{$(1.7 \pm 0.2) \times 10^{-3}$} & \multicolumn{2}{c}{$(1.7 \pm 0.2) \times 10^{-3}$} \\
     MSSI                            & \multicolumn{2}{c}{$(1.4 \pm 1.4) \times 10^{-4}$} & \multicolumn{2}{c}{$(1.3 \pm 1.0) \times 10^{-4}$} \\
     Detector NRs                    & \multicolumn{2}{c}{--} & \multicolumn{2}{c}{$[0, 1.3]$}$\,\,\,\,\,\,\,\,\,\,\,\,\,\,\,\,$ \\
     $\mathcal{L}_{10}^s$ (1000 GeV/c$^2$)                & \multicolumn{2}{c}{--} & \multicolumn{2}{c}{$\left(3.1^{+4.2}_{-2.1}\right) \times 10^{-2}$} \\
     \hline
     $\lambda_{\text{DN}}$           & \multicolumn{2}{c}{$(8.7 \pm 0.2) \times 10^{-1}$} & \multicolumn{2}{c}{$(8.7 \pm 0.1) \times 10^{-1}$} \\
     $\lambda_{\text{PN}}$           & \multicolumn{2}{c}{$(5.0 \pm 1.0) \times 10^{-2}$} & \multicolumn{2}{c}{$(5.0 \pm 0.7) \times 10^{-2}$} \\ \hline
     \hline
     Total counts                    & \multicolumn{2}{c}{--} & 50.4 & 2.0 \\
     \hline
     Observed counts                 & \multicolumn{2}{c}{--} & \multicolumn{2}{c}{55} $\,\,\,\,\,\,\,\,\,\,\,\,\,\,\,$\\
     \hline \hline
    \end{tabular}
    \label{tab:delayed_sample_fit_table} 
\end{table}

\subsection{Details of the Detector Response modeling}

During development of the detector response model, NEST v2.4.0~\cite{nest_v2_4_0} was found not to reproduce the measured ER-band widths across the full calibration energy range, due to inaccurate modelling of recombination fluctuations. In NEST v2.4.0, $\sigma_p$ is a parameter controlling the component of recombination fluctuations that depends on the  number of ions~\cite{NEST:paper_2023}, modeled as a single skewed Gaussian. The discrepancy was resolved in NEST~v2.4.5~\cite{nest_v2_4_5}, with
the final model parameterizing $\sigma_p$ as the sum of two skewed Gaussian functions of the number of quanta, $N_q$:
\begin{equation}
    \sigma_p(x) = \sum_{i=1}^2 \left[A_i\left[1 + \mathrm{erf}\left(\frac{\alpha_i (x - \mu_i) }{\sqrt{2}\sigma_i}\right)\right]\exp\left(-\frac{(x-\mu_i)^2}{2\sigma_i^2}\right)\right] 
\end{equation}
where $x = \log_{10}(N_q)$, and $A_i$, $\alpha_i$, $\mu_i$, and $\sigma_i$ are the parameters of the skew-Gaussians. This modification provides good agreement with the measured ER-band widths across the calibrated range. Tables~\ref{tab:eft2024_er_mean_yield} and~\ref{tab:eft2024_er_double_gaussian_parameters} show the NEST parameters used in the final ER model. 

The NEST v2.4.0 model for nuclear recoils below 75 keV was well calibrated with DD neutrons for Ref.~\cite{LZ:2025lowmass}, but including the higher-energy AmBe calibration data revealed a tension between the NR charge yield parameters preferred by the DD and AmBe regions, preventing a consistent parameter set across the full energy range. In NEST, the NR charge yields are governed by a power law with exponent $p$. To resolve the tension, NEST v2.4.5 introduces a break in the power law such that  $p=0.5$ below a threshold energy $E_0$ and follows a logarithmic energy dependence above $E_0$, 
\begin{equation}
    p(E)=0.5 + a \ln\!\left[1 + b\,(E - E_0)\right].
\end{equation}
Here, $a$ controls the overall suppression strength, and $b$ controls how the suppression develops with recoil energy.
This treatment allows the low- and high-energy responses to be described within a unified NR model.
Previous work identified an abrupt change in the electronic stopping-power behaviour of xenon nuclear recoils~\cite{Aprile:2006} and a preference for a larger value of the power-law exponent at high recoil energies~\cite{Lenardo:2015}, supporting the need for such a modification. Table~\ref{tab:eft2024_nr_yield_parameters} shows the NEST parameters used in the final NR model.

\begin{table*}[ht]
    \centering
    \caption{
       Model parameters for the NEST ER mean
        charge-yield model.
    }
    \begin{tabular}{lcc}
        \hline\hline
        Parameter & {NEST Default at 97 V/cm} & {Value in Analysis} \\
        \hline
        $m_{1}$ [$\mathrm{keV}^{-1}$] & 14.6 & 12.31 \\
        $m_{2}$ [$\mathrm{keV}^{-1}$] & 77.3 & 84.91 \\
        $m_{3}$ [$\mathrm{keV}$]      & 0.8  & 0.5707 \\
        $m_{4}$                       & 2.1  & 2.804 \\
        $m_{5}$ [$\mathrm{keV}^{-1}$] & 19.3 & 34.04 \\
        $m_{6}$ [$\mathrm{keV}^{-1}$] & 0    & 0 \\
        $m_{7}$ [$\mathrm{keV}$]      & 78.0 & 82.57 \\
        $m_{8}$                       & 4.3  & 4.557 \\
        $m_{9}$                       & 0.33 & 0.2207 \\
        $m_{10}$                      & 0.08 & 0.1272 \\
        \hline\hline
    \end{tabular}
    \label{tab:eft2024_er_mean_yield}
\end{table*}

\begin{table*}[ht]
    \centering
    \caption{Tuned parameters for the double skewed-Gaussian ER fluctuation model.}
    \begin{tabular}{lc}
        \hline\hline
        Parameter & Value in Analysis \\
        \hline
        $A_1$      & 0.03174 \\
        $\mu_1$    & 2.588 \\
        $\sigma_1$ & 0.3627 \\
        $\alpha_1$ & -0.3583 \\
        \hline
        $A_2$      & 0.06211 \\
        $\mu_2$    & 5.2 \\
        $\sigma_2$ & 1.359 \\
        $\alpha_2$ & -2.599 \\
        \hline\hline
    \end{tabular}
    \label{tab:eft2024_er_double_gaussian_parameters}
\end{table*}

\begin{table*}[ht]
    \centering
    \caption{
        Model parameters for the NEST NR mean light- and charge-yield model.
        The parameters $f_1$ and $f_2$ control the asymmetric low-energy
        turn-on of the charge and light yields, respectively. Parameters $a$, $b$, and $E_0$ control the energy-dependent modification of $p$.
    }
    \begin{tabular}{lcc}
        \hline\hline
        Parameter & NEST Default & Value in Analysis \\
        \hline
        $\alpha$ [$\mathrm{keV}^{1/\beta}$] & 11.0 & 11.2 \\
        $\beta$ & 1.1 & 1.1 \\
        $\gamma$ [$(\mathrm{V/cm})^{1/\delta}$] & 0.048 & 0.052 \\
        $\delta$ & -0.0533 & -0.0533 \\
        $\epsilon$ [$\mathrm{keV}$] & 12.6 & 10.8 \\
        $\zeta$ [$\mathrm{keV}$] & 0.30 & 0.53 \\
        $\eta$ & 2.0 & 1.4 \\
        $\theta$ [$\mathrm{keV}$] & 0.30 & 0.31 \\
        $\iota$ & 2.0 & 2.5 \\
        $p$ & 0.50 & 0.50 \\
        \hline
        $f_1$ & 1.0 & 1.39 \\
        $f_2$ & 1.0 & 1.74 \\
                \hline
        $a$ & NA & 0.0230 \\
        $b$ & NA &  0.0289 \\
        $E_0$ [$\mathrm{keV}$] & NA & 74.7 \\
        \hline\hline
    \end{tabular}
    \label{tab:eft2024_nr_yield_parameters}
\end{table*}

\subsection{Waveform Analysis}
The shapes of the S1 and S2 pulses can provide additional information about an event. Scintillation light in LXe is produced from the decay of a metastable dimer Xe$^*_2$, which exists in either a singlet or triplet molecular state with lifetimes of $2-4$~ns and $21-28$~ns, respectively.
ER and NR interactions populate the singlet and triplet states with different ratios, with NRs having a higher proportion of singlet decays and therefore a faster S1, providing some discrimination power between interaction types~\cite{PSD:Theory,PSD:time1,PSD:time2,PSD:time3,XMASS:2018hil,LZ:PSD_2026}.
In a TPC as large as LZ, the S1 shape also has a position dependence, with photons produced further from PMTs experiencing a broader distribution of propagation times prior to collection.
Unfortunately, the S1 pulse timing has limited discrimination power on an event-by-event basis for events with S$1c\sim550$~phd and the event of interest is generally compatible with both ER and NR populations. Existing NR calibration data around 250~keV are limited, and there are no such calibration events within a few cm of the event of interest. We find that AmBe calibration NRs with approximately 150~keV are not inconsistent with an ER template built from $^{131\text{m}}$Xe calibration data near the position of the event of interest, showing the lack of event-by-event discrimination power.

The pulse timing provides more discrimination regarding the $z$-position of the event. Both ER and NR events near the bottom of the detector have fast rise times because light can travel more directly to the bottom PMT array.  Examination of waveforms for a sample of events from below the FV near the cathode, where most RFR MSSI events would be found, finds shorter rise times than all samples at the same depth as the event of interest. The waveform of the event of interest is thus less consistent with that of a RFR MSSI than with a NR at the reconstructed position.

Pulse shapes can also inform the classification of an event as a single scatter event from near the reconstructed position. Examination of the S2 shape of the event of interest finds consistency with a template for single-site interactions around the same location generated from a selection of alpha decays. The partitioning of S1 between the top and bottom PMT array (also known as the ``top-bottom asymmetry'') for the event of interest is also consistent with a selection of events from the same $z$. Examination of the S1 hit pattern across all the individual PMTs in the PMT arrays finds consistency at the 2$\sigma$ level in $z$ and 1$\sigma$ level in $(x,y)$ with a selection of SS ER events of similar S1 size and location as the event of interest. Analysis of the S1 pattern shows inconsistency with an RFR MSSI template, but this analysis could not distinguish between a single scatter at the reconstructed position of the event of interest and a wall MSSI template.  

\subsection{Accidental backgrounds}
The accidentals model is constructed in similar fashion to Refs.~\cite{LZ:SR3_WS2024,LZ:2025lowmass}. A high statistics synthetic dataset is generated from isolated S1 and S2 pulses that pass a minimal set of data quality selections. These S1 and S2 pulses are paired at the waveform level to generate a sample of accidental coincidence events, containing pairs of pulses that have either physical drift times (PDTs) or unphysical drift times (UDTs) depending on whether the apparent drift time is shorter or longer than the drift time for events at the cathode. 

\begin{figure}[!h]
    \includegraphics[width=0.45\columnwidth]{\plotfolder 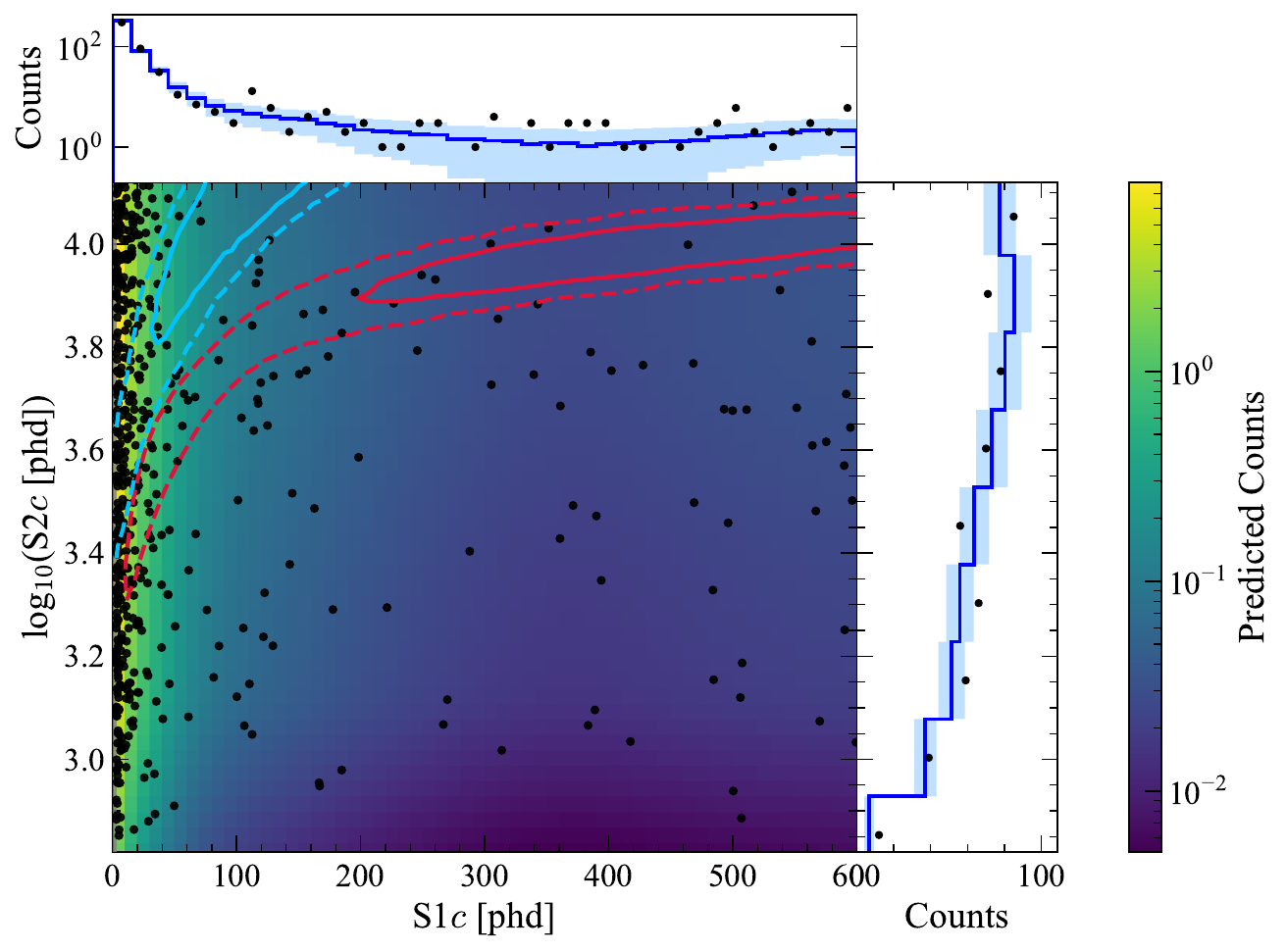}
    \includegraphics[width=0.45\columnwidth]{\plotfolder 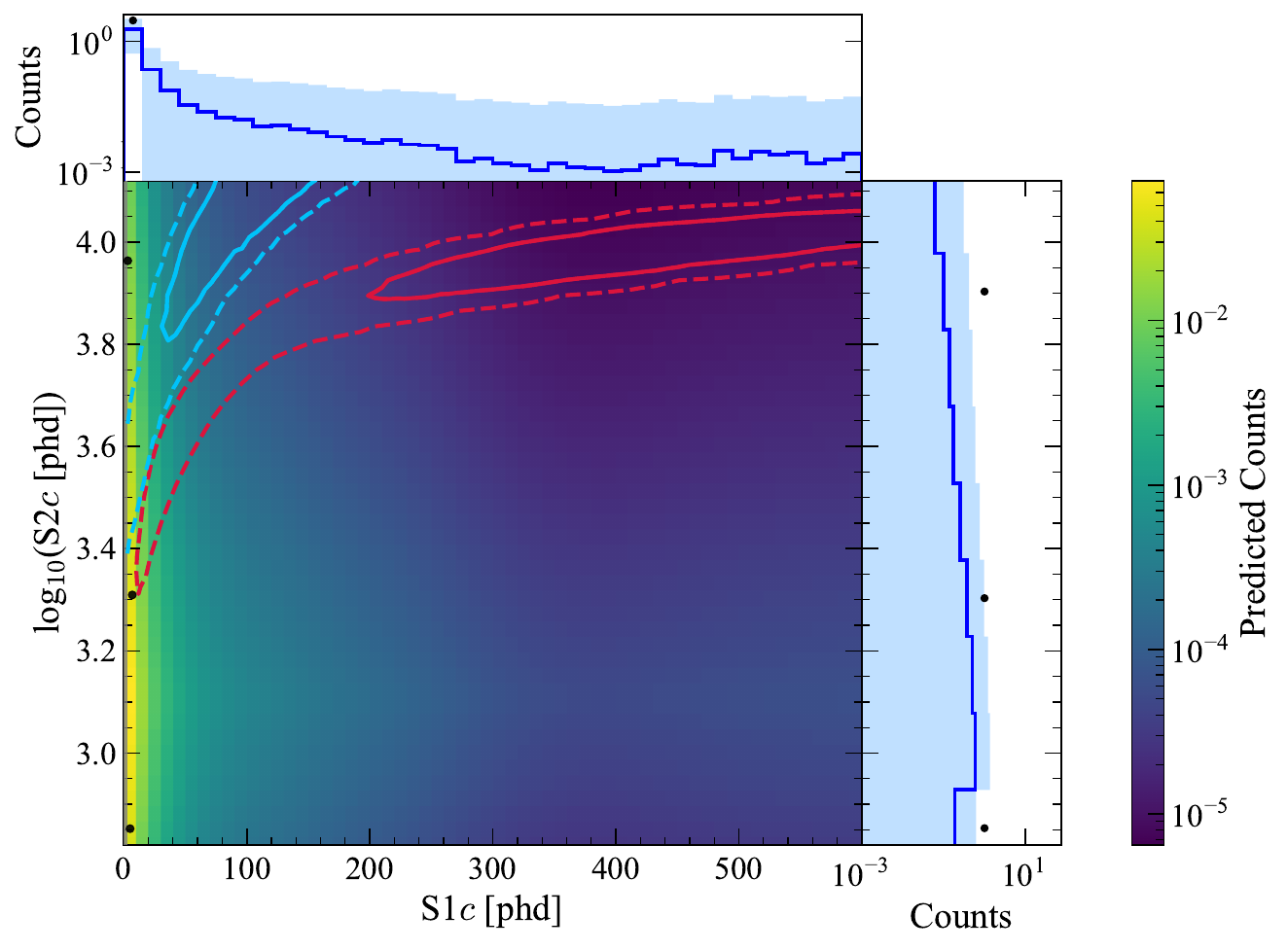}
    \caption{\label{fig:accidentals} Probability density function in \{S1$c$, $\log_{10}(\mathrm{S}2c)$\} space for the accidentals model compared to unphysical drift time events in the data: (left) with initial loose data quality cuts when the model is normalized to the data and (right) after all cuts targeting the accidental background. 1D projections of each axis are also shown with the blue band representing the combined systematic and statistical uncertainty on the model. The blue and red regions represent the continuous ER background and the best fit 1000~GeV/c$^2$ $\mathcal{L}_{10}^s$ WIMP model.
    After all cuts, the accidentals model predicts $2.4\pm0.2$ events in the UDT dataset, with 3 events observed. 
    }
\end{figure}

The model is validated by comparing to the UDT population in data, as these are by definition accidental pairings of S1 and S2 pulses. In the validation, the synthetic dataset of UDTs is normalized to the real UDT dataset before applying pulse-based selections that specifically target accidental populations. For example, one pulse-based selection ensures that the S2 pulse width matches the inferred drift time of the event. Model and UDT agreement is checked as each cut is applied, and all comparisons are within 20\%. After all cuts are applied, the normalized synthetic model predicts $2.4\pm0.2$ UDT events, with 3 remaining events in the UDT dataset. The left panel of \autoref{fig:accidentals} shows the \{S1$c$,~$\log_{10}(\mathrm{S}2c)$\} distribution of the UDT model and UDT population at the normalization step, along with 1D projections. The right panel shows the same distributions after all selections are applied.  The good agreement is taken to indicate that the PDT component is also well-modeled. 

\subsection{Multiple-scintillation single-ionization (MSSI) events}
The MSSI background is of interest as it can populate regions of \{S1$c$, $\log_{10}(\mathrm{S}2c$\} space below the ER band, but generally at reconstructed energies above those explored in Refs.~\cite{LZ:SR1WS_2022, LZ:SR3_WS2024}. The basic topology is that the S1 and S2 of a typical ER are combined with an S1 from a second scatter in a charge dead region in the RFR or near the wall, moving an ER event horizontally in the \{S1$c$, $\log_{10}(\mathrm{S}2c$\} plane. Various combinations of scatters can result in MSSI, including:
\begin{itemize}
    \item{$\beta$ decay in the FV with an associated $\gamma$-ray depositing energy in a charge dead region before leaving the detector}
    \item{The inverse, where a $\beta$ decays in a charge dead region and the $\gamma$-ray scatters or photoabsorbs in the FV}
    \item{Compton scattering of a $\gamma$-ray emitted from detector components or internal radioactivity, with one scatter in the FV and one or more scatters in a charge dead region}
    \item{Decays of radon daughters or other isotopes that plate out on the cathode or TPC walls, such that a $\beta$, $\alpha$, or NR is degraded by the relevant plateout surface with associated $\gamma$-rays interacting in the FV and/or RFR.}
\end{itemize}

As described in the main text, the MSSI model is derived from simulations of detector materials  and $^{214}$Pb decays distributed throughout the LXe. Other distributed radioactive isotopes, including $^{124}$Xe, $^{127}$Xe, $^{133}$Xe, $^{125}$I were simulated and shown to be subdominant. Low energy ($\lesssim 1.5$~MeV) $\gamma$-rays in general do not penetrate into the analysis volumes without depositing more than 100~keV in the LXe. Radon plate-out on the cathode was studied in Ref.~\cite{LZ:2026hpq} and found to have similar levels of activity as radioactivity in the cathode wires themselves, which are already included in the detector materials. A concern in radon daughter plate-out is high energy $\gamma$-rays associated with $^{214}$Bi, but a significant fraction of those will be observed with $^{214}$Po alpha decay. These contributions were therefore not included in the MSSI model.

\begin{figure}[!h]
    \includegraphics[width=\columnwidth]{\plotfolder 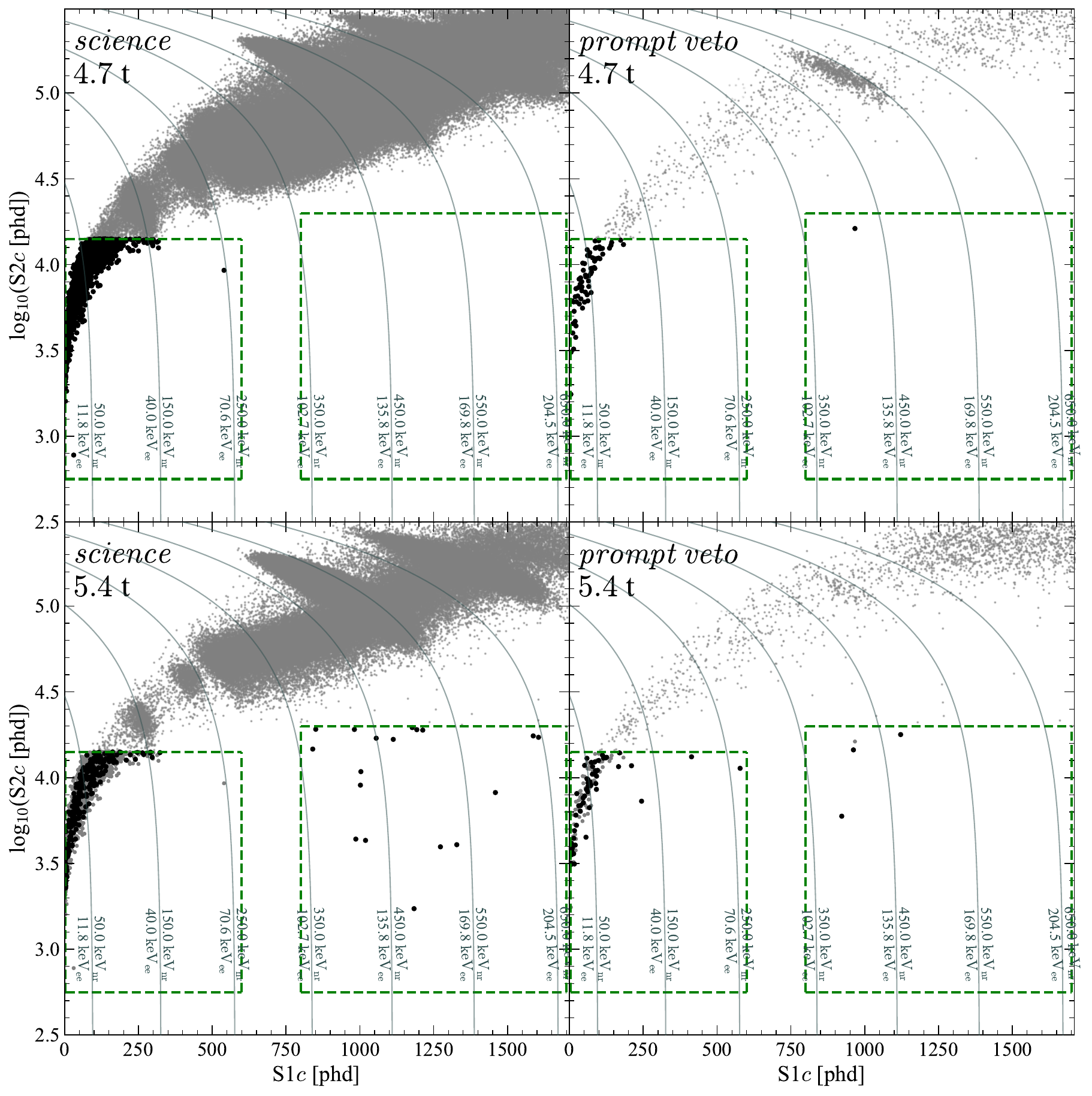}
    \caption{\label{fig:MSSI_samples} Samples used for validation of the MSSI model in \{S1c, log$_{10}$(S2c)\} space: the 4.7~tonne FV used in this analysis (top row), an increased analysis volume of 5.4~tonne (bottom row), \emph{science} sample (left column), and \emph{prompt veto} sample (right column). 
    The two green dashed boxes show the analysis regions for the comparison, with the WS~ROI for the analysis at lower S1 and the HE~SB at higher S1. Counts that appear in the green boxes in the 4.7~tonne FV are shown in grey in the larger 5.4~tonne volume. 
    The observed counts in each sample are compared to the model predictions in~\autoref{tab:MSSI_comp}. 
    }
\end{figure}

\begin{figure}[!h]
    \includegraphics[width=0.45\columnwidth]{\plotfolder 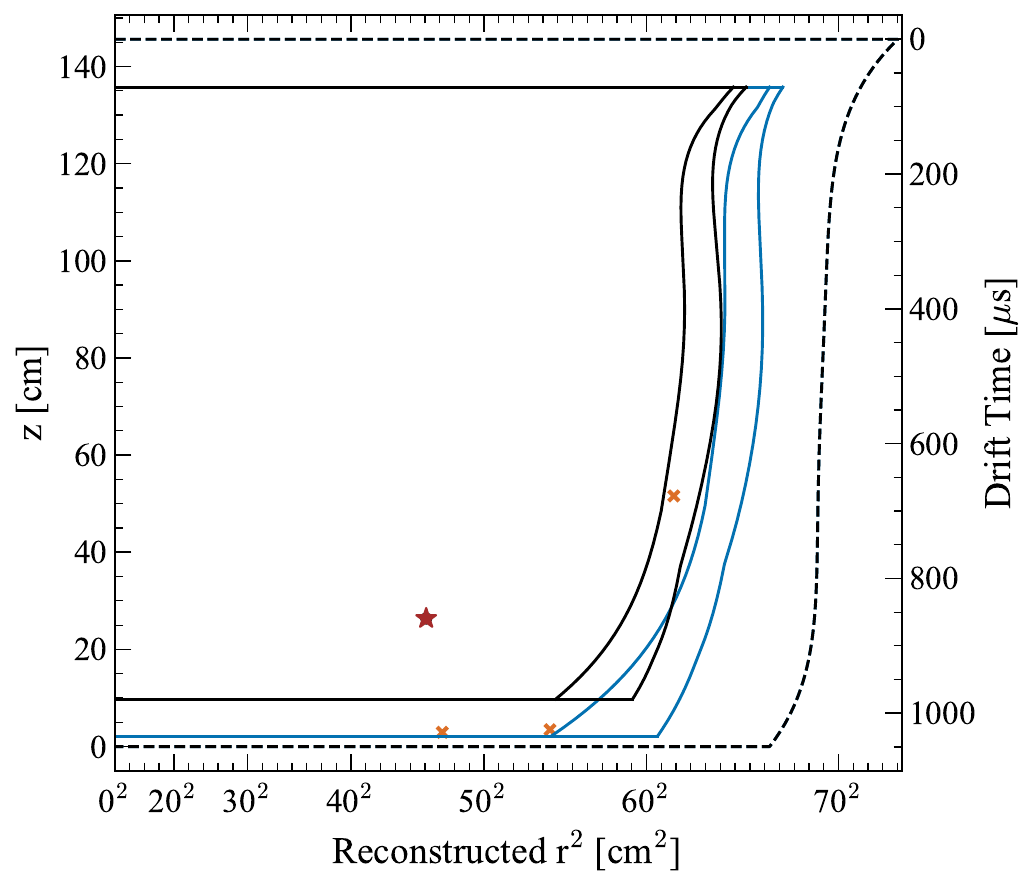}
    \includegraphics[width=0.45\columnwidth]{\plotfolder 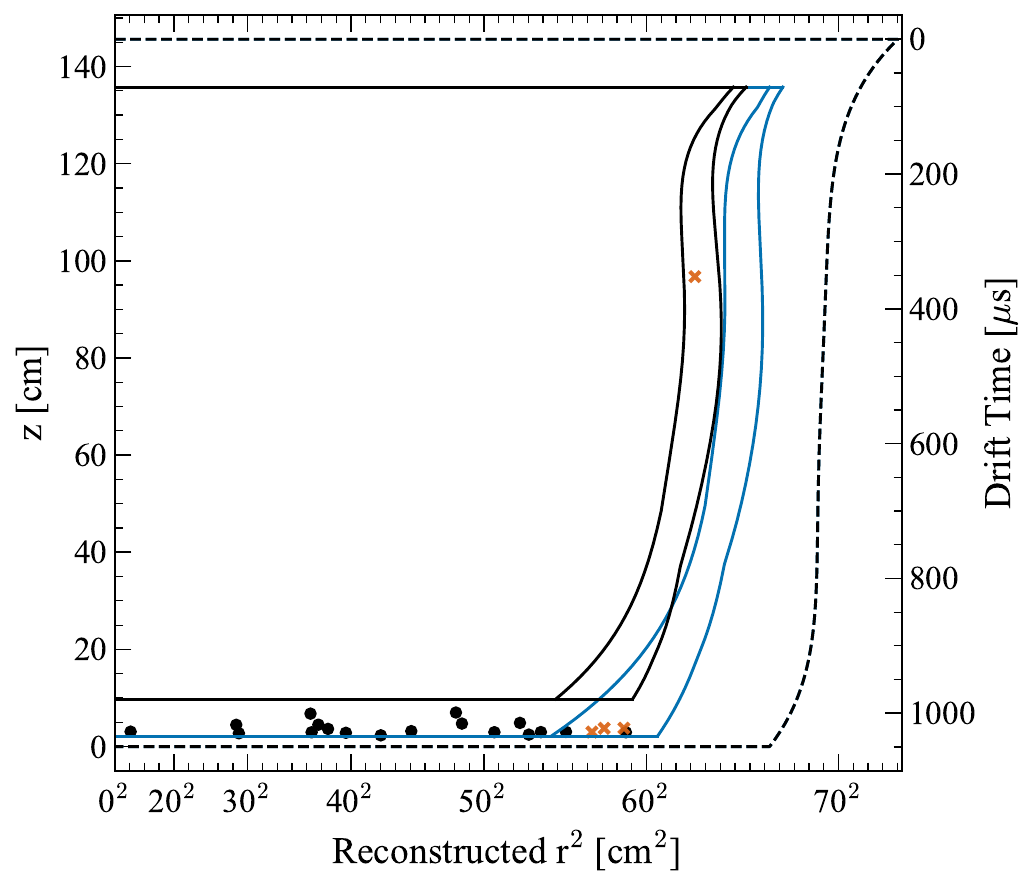}
    \caption{Reconstructed $r^2-z$ projections for the two analysis regions: WS~ROI of the main analysis (left) and the high energy sideband (HE~SB) used to validate the MSSI model (right). 
    The \emph{science} and \emph{prompt veto} samples are denoted by black circles or orange crosses, respectively. 
    In the WS~ROI, the events shown are 5$\sigma$ or greater from the ER band, with the event of interest shown as the brown star.
    The dashed line indicates the reconstructed wall, averaged over azimuth, indicating the active volume.
    The solid lines indicate the minimum and maximum radial extent of the 4.7~t (black) and 5.4~t (blue) analysis volumes.
    }.
    \label{fig:MSSI_FVs}
\end{figure}

\begin{figure}[!h]
    \includegraphics[width=\columnwidth]{\plotfolder 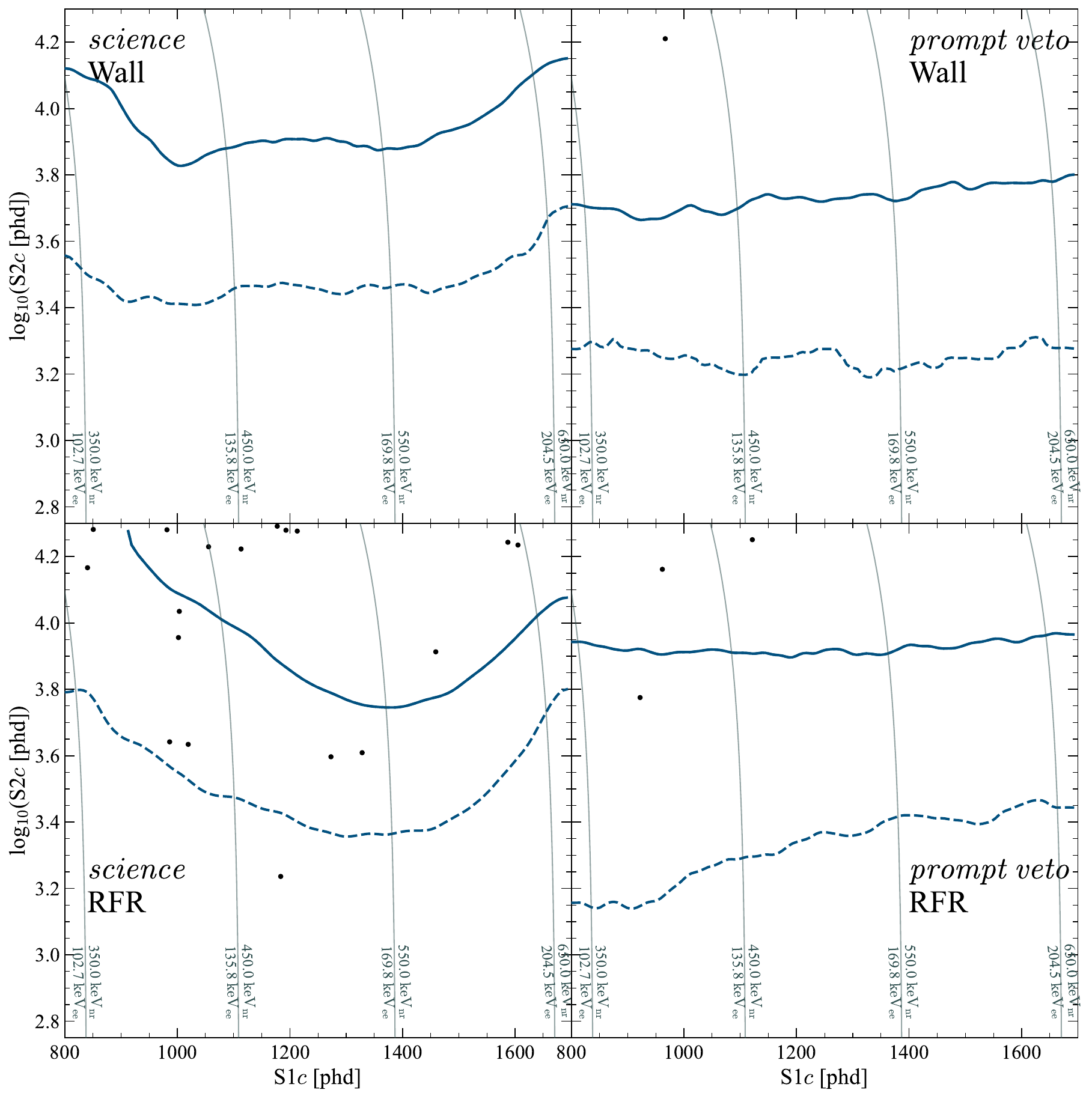}
    \caption{\label{fig:MSSI_contours}
    Data in the high energy sideband (HE~SB) used for validation of the MSSI model in \{S1c, log$_{10}$(S2c)\} space for the 5.4~tonne LXe analysis volume, split up by     \emph{science} sample (left column), \emph{prompt veto} sample (right column), wall MSSI (top row), and RFR MSSI (bottom row).
    The contours are the 68\% (solid) and 95\% (dashed) containment regions for MSSI of that type in the relevant sample.
    }
\end{figure}

The MSSI simulation is validated with a 5.4 tonne analysis volume, events in a higher energy sideband to the WS~ROI (HE~SB: 800 $<$ S1$_c$ $<$ 1700~phd, $10^{2.75}$ $<$ S2$c$ $<$ $10^{4.3}$~phd), and the \emph{prompt veto} sideband.  The 5.4 tonne analysis volume uses the same $z$ boundaries as Ref.~\cite{LZ:SR3_WS2024} with a radial boundary defined by requiring an expected wall background count of less than $(1.0 \pm 0.5)\times10^{-2}$ in the $3-600$~phd WS~ROI; the radial boundary in Ref.~\cite{LZ:SR3_WS2024} was defined by the same criterion over the $3-80$~phd ROI used in that analysis. 
Figure~\ref{fig:MSSI_samples} shows the \emph{science} and \emph{prompt veto} samples for both analysis volumes in \{S1c, log$_{10}$(S2c)\} space, and Figure~\ref{fig:MSSI_FVs} shows $r^2-z$ projections for the WS ROI and the HE SB. 
\autoref{tab:MSSI_comp} shows the counts in the various subsamples used to validate the MSSI model. The veto efficiency $\lambda_{MSSI}$ is much lower in the HE~SB of the 5.4-tonne analysis volume than the $94\pm2$\% reported for the main analysis, because the primary source of MSSI events in that sample is $^{214}$Pb RFR MSSI where the gamma is fully contained in the RFR.
The model shows good agreement with each subsample, including the division between RFR and wall events; a binned Poisson likelihood chi-square test using the numbers in \autoref{tab:MSSI_comp} returns a p-value of 0.7~\cite{BAKER1984437}.
Figure~\ref{fig:MSSI_contours} further splits up the HE~SB of the 5.4~tonne analysis volume into Wall and RFR components in the \emph{science} and \emph{prompt veto} samples, with contours showing the 68\% (solid) and 95\% (dashed) containment of the simulated MSSI model.
The contours show broad agreement with the data, albeit with limited statistics. 

In the HE~SB, the dominant component of the \emph{science} sample is predicted to be RFR events from $^{214}$Pb MSSI. 
As a cross check, we deploy the radon tag~\cite{LZ:2025xxf}; 12 events occur in radon-tag-active periods, and 7 events are tagged, consistent with the measured radon tagging efficiency of 60\%. 
In the same HE~SB, the \emph{veto} sample is expected to be dominated by detector radioactivity; the wall event and two of the RFR events are in radon-tag-active periods, with one of the RFR events radon tagged.  
These findings support the overall compatibility of the model with the sideband data.  

Under the hypothesis that the event of interest is MSSI consisting of two ER deposits where the first scatter is at the location of the event with S1 and S2 observed and the second scatter in a charge dead region with only S1 observed, the two scatters can be reconstructed. For an observed S2$c$ of 9268~phd, the expected energy of the first scatter is $12\pm2$~keV with an expected S1$c$ of $69\pm17$~phd, leaving approximately 471 phd for the S1$c$ of the second scatter. 
Interpreted as a wall MSSI event (assuming the same $g_1$ as the bulk of the LXe), the second scatter would be $77\pm7$~keV. Interpreted as an RFR MSSI event (where the second scatter is in the high field of the RFR with suppressed recombination), the second scatter would be $204^{+65}_{-38}$~keV.  
While we use event positions to determine whether or not an event is in the FV, we do not use the position explicitly in the statistical inference. We note that it is unlikely for any MSSI category to include a scatter with only $\sim$12~keV of energy deposition more than 20 cm from the boundaries of the TPC. Compton scattering of a high energy $\gamma$-ray requires a very shallow scattering angle to deposit so little energy, and thus the $\gamma$-ray must traverse more than 60~cm through LXe without a second interaction. A low energy $\gamma$-ray is very unlikely to penetrate far enough to photoabsorb.

\begin{table}[htbp]
  \centering
  \caption{Comparison of simulated and observed MSSI counts. The ``WS~ROI'' is the WIMP search region of interest used in the main analysis with ``HE~SB'' referring to the higher energy sideband. ``RFR'' events fall close to the cathode and ``wall'' events are higher in the TPC along the TPC walls. Two analysis volumes are shown, the 4.7 tonne fiducial volume used in the main analysis and a 5.4 tonne analysis volume similar to that used in previous WIMP searches~\cite{LZ:SR3_WS2024}; in this table, the analysis volumes are disjoint, with the larger one including only counts outside of the nested inner FV. A binned Poisson likelihood chi-square test excluding the data in the main analysis (WS~ROI, 4.7~tonne FV) finds a p-value of 0.7, suggesting compatibility. }
  \setlength{\tabcolsep}{6pt} 
  \begin{tabular}{lcccccc}
    \hline\hline
    \multirow{2}{*}{Region} & MSSI &  Analysis Vol. & {Simulated}    & Observed   & Simulated  &Observed   \\ 
     & type & (tonne) & \emph{science} &\emph{science} & \emph{prompt} & \emph{prompt} \\   \hline 
    WS~ROI &Wall & 4.7 & 0.0048   & --  & 0.09 & -- \\
    WS~ROI &RFR & 4.7  & 0.0001  & --  & 0.10 & -- \\ \hline
    WS~ROI &Wall & 5.4 & 0.03   & 0  & 0.1 & 1 \\
    WS~ROI &RFR & 5.4  & 0.003  & 0  & 1.9 & 2 \\
     HE~SB &Wall & 4.7   & 0.1   & 0  & 0.3 & 1 \\
    HE~SB &RFR & 4.7    & 0.5   & 0  & 0.5 & 0 \\
    HE~SB &Wall & 5.4   & 0.5   & 0  & 2.2 & 0 \\
    HE~SB &RFR & 5.4    & 21.5  & 18 & 5.2 & 3 \\   \hline\hline
  \end{tabular}
  \label{tab:MSSI_comp}
\end{table}

\clearpage

\subsection{Neutrons}

The dominant neutron component comes from ($\alpha$,n) reactions in detector materials. 
Spontaneous fission neutrons are included in the statistical inference although the rate is up to two orders of magnitude lower than ($\alpha$,n): the high neutron and $\gamma$-ray multiplicity of spontaneous fission leads to a high probability of multiple interactions in the LZ detectors~\cite{LZ:simulations_2021}. 
As a cross-check, multiple scatter neutrons were analyzed to provide a secondary estimate for the number of SS neutrons in the WS~ROI, predicting 0.02$\pm$0.02.

Muon-induced cascades can produce neutrons of up to GeV energies that are likely to be aligned with the muon direction.  
Simulations of 2.1$\times10^{9}$ atmospheric muon events were generated using MUSUN~\cite{Kudryavtsev:2008qh}, equivalent to $1200$ years of LZ exposure, with no signal-like deposits seen.
The muon simulations are validated against GEANT4 and FLUKA neutron yields~\cite{ARAUJO2005398, Pec:2023yic}, with an estimated uncertainty of a factor of 2. 
Additional simulations of neutrons with energies ranging from 50~MeV -- 1~GeV produced in rock were carried out to evaluate the tagging efficiency of the veto system for the case when neither a muon nor any other secondary particle trigger any detector. 
This tagging efficiency was found to be $80.0\pm4.0$\%. 
We set a 90\% confidence level upper limit on the number of predicted SS events of any energy from muon-induced neutrons of $4.6 \times 10^{-4}$, and we do not include muon-induced neutrons in the background model.

\subsection{Interpretation of $\mathcal{O}_1$ as a cross-section}
\begin{figure}[!h]
    \includegraphics[width=0.5\columnwidth]{\plotfolder 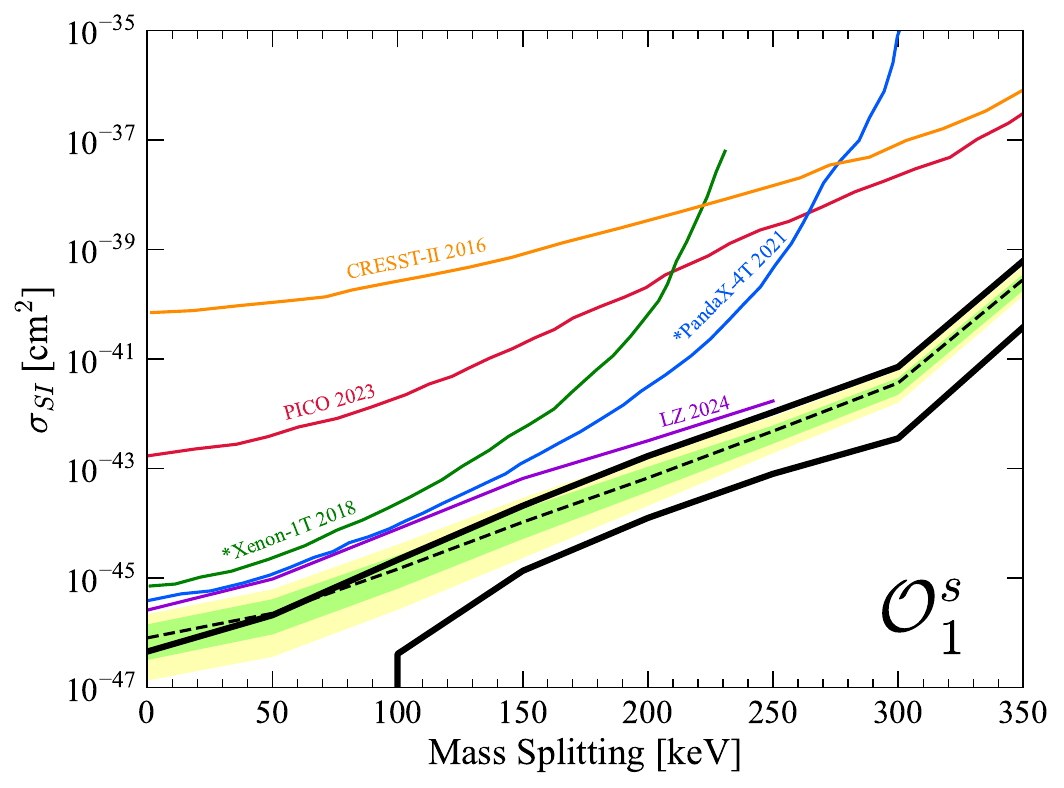}
    \caption{\label{fig:inelastic_as_si}
    Two-sided 90\% confidence-level intervals (solid-black) on inelastic DM-nucleon scattering cross sections assuming scalar normalization.
    The black dotted line is the median sensitivity, and the green and yellow bands contain 68\% and 95\% of expected upper limits given the best-fit background model.
    Previous limits are shown for CRESST-II (orange)~\cite{inelasticlimits:creest_2020,Bramante2016:inelasticDM}, PICO (red)~\cite{PICO:InelasticDM_2023}, Xenon-1T (green)~\cite{inelasticlimits:xenon1t}, PandaX-4T (blue)~\cite{PandaX4T:SI2023}, and LZ (violet)~\cite{LZ:SR1_NREFT_2023}.
    Limits with a star indicate the limit was derived by the PICO collaboration~\cite{PICO:InelasticDM_2023}. 
    }
\end{figure}

This analysis was performed in the isoscalar and isovector basis.
To enable comparison to SI cross-sections, the constraints set on $\mathcal{O}_1^s$ can be recast through 
\begin{equation}
    (c_1^s \times m_\nu^2)^2 = \sigma^N_{SI}\frac{\pi \cdot m_\nu^4}{\mu^2},
\end{equation}
where $m_\nu$ is the Higgs vacuum expectation value and $\mu$ is the reduced mass of the WIMP-nucleon system.
Figure~\ref{fig:inelastic_as_si} shows the recasting of $\mathcal{O}_1^s$ assuming scalar coupling.
These limits can additionally be converted to limits on the cross-section for vector coupling by multiplying by a factor $(A/((A-Z)-(1-4\sin^2\theta_w)Z))^2\simeq 3.2$ (for Xe).

\subsection{Local significance}
Table~\ref{tab:lagrangian_significance} contains the local significance of the Lagrangian models tested, and Table~\ref{tab:operator_significance} shows the same for the $\mathcal{O}_i$ models.
A minimum of 30,000 toy distributions were generated for each case.
The toys are used to determine the $p_0$ value of the fit to our data, from Eq.~6 of Ref.~\cite{DM_parameters:BAXTER2021_Conventions}.  The value in the table is the number of normal-distribution standard deviations (one-sided) that correspond to the observed $p_0$ for each model and WIMP mass.

\begin{table*}[ht]
    \centering
    \caption{Local significance for Lagrangian models tested}
    \begin{tabular}{l|lllllllllllll}
    \hline
    \multirow{2}{*}{L$_i$} & \multicolumn{13}{c}{WIMP Mass (GeV/c$^2$)} \\
    & 10 & 12 & 14 & 17 & 21 & 30 & 40 & 50 & 100 & 200 & 400 & 1000 & 4000 \\
    
    \hline
    L$^s_1$ & 0.0 & 0.0 & 0.0 & 0.0 & 0.0 & 0.0 & 0.0 & 0.0 & 0.0 & 0.0 & 0.0 & 0.0 & 0.0 \\
    L$^v_1$ & 0.0 & 0.0 & 0.0 & 0.0 & 0.0 & 0.0 & 0.0 & 0.0 & 0.0 & 0.0 & 0.3 & 1.3 & 1.2 \\
    L$^s_2$ & 0.0 & 0.0 & 0.0 & 0.0 & 0.0 & 0.0 & 0.0 & 0.0 & 2.4 & 2.8 & 2.9 & 3.0 & 3.1 \\
    L$^v_2$ & 0.0 & 0.0 & 0.0 & 0.0 & 0.0 & 0.0 & 0.0 & 0.0 & 2.4 & 2.5 & 3.0 & 3.1 & 3.1 \\
    L$^s_3$ & 0.0 & 0.0 & 0.0 & 0.0 & 0.0 & 0.0 & 0.0 & 0.0 & 0.0 & 1.1 & 1.5 & 1.7 & 1.8 \\
    L$^v_3$ & 0.0 & 0.0 & 0.0 & 0.0 & 0.0 & 0.0 & 0.0 & 0.0 & 0.2 & 2.2 & 2.5 & 2.6 & 2.7 \\
    L$^s_4$ & 0.0 & 0.0 & 0.0 & 0.0 & 0.0 & 0.0 & 0.0 & 0.1 & 2.6 & 3.1 & 3.1 & 3.1 & 3.0 \\
    L$^v_4$ & 0.0 & 0.0 & 0.0 & 0.0 & 0.0 & 0.0 & 0.0 & 0.1 & 2.7 & 3.1 & 3.2 & 3.1 & 3.1 \\
    L$^s_5$ & 0.0 & 0.0 & 0.0 & 0.0 & 0.0 & 0.0 & 0.0 & 0.0 & 0.0 & 0.0 & 0.0 & 0.0 & 0.0 \\
    L$^v_5$ & 0.0 & 0.0 & 0.0 & 0.0 & 0.0 & 0.0 & 0.0 & 0.0 & 0.0 & 0.0 & 0.7 & 1.3 & 1.3 \\
    L$^s_6$ & 0.0 & 0.0 & 0.0 & 0.0 & 0.0 & 0.0 & 0.0 & 0.0 & 1.1 & 2.2 & 2.6 & 2.6 & 2.6 \\
    L$^v_6$ & 0.0 & 0.0 & 0.0 & 0.0 & 0.0 & 0.0 & 0.1 & 0.1 & 2.9 & 3.1 & 3.2 & 3.3 & 3.3 \\
    L$^s_7$ & 0.0 & 0.0 & 0.0 & 0.0 & 0.0 & 0.0 & 0.0 & 0.0 & 1.6 & 2.5 & 2.7 & 2.7 & 2.7 \\
    L$^v_7$ & 0.0 & 0.0 & 0.0 & 0.0 & 0.0 & 0.0 & 0.0 & 0.0 & 1.6 & 2.3 & 2.7 & 2.7 & 2.8 \\
    L$^s_8$ & 0.0 & 0.0 & 0.0 & 0.0 & 0.0 & 0.0 & 0.0 & 0.0 & 2.3 & 2.8 & 2.9 & 3.0 & 3.0 \\
    L$^v_8$ & 0.0 & 0.0 & 0.0 & 0.0 & 0.0 & 0.0 & 0.0 & 0.0 & 2.4 & 2.9 & 2.9 & 3.0 & 3.0 \\
    L$^s_9$ & 0.0 & 0.0 & 0.0 & 0.0 & 0.0 & 0.0 & 0.0 & 0.0 & 2.1 & 2.9 & 2.9 & 3.0 & 3.1 \\
    L$^v_9$ & 0.0 & 0.0 & 0.0 & 0.0 & 0.0 & 0.0 & 0.0 & 0.0 & 2.7 & 3.0 & 3.1 & 3.2 & 3.1 \\
    L$^s_{10}$ & 0.0 & 0.0 & 0.0 & 0.0 & 0.0 & 0.0 & 0.0 & 0.0 & 2.9 & 3.1 & 3.4 & 3.4 & 3.4 \\
    L$^v_{10}$ & 0.0 & 0.0 & 0.0 & 0.0 & 0.0 & 0.0 & 0.0 & 0.0 & 3.0 & 3.2 & 3.3 & 3.4 & 3.4 \\
    L$^s_{11}$ & 0.0 & 0.0 & 0.0 & 0.0 & 0.0 & 0.0 & 0.0 & 0.0 & 2.5 & 3.0 & 3.1 & 3.2 & 3.2 \\
    L$^v_{11}$ & 0.0 & 0.0 & 0.0 & 0.0 & 0.0 & 0.0 & 0.0 & 0.0 & 2.5 & 2.9 & 3.1 & 3.2 & 3.2 \\
    L$^s_{12}$ & 0.0 & 0.0 & 0.0 & 0.0 & 0.0 & 0.0 & 0.0 & 0.0 & 2.4 & 3.1 & 3.2 & 3.2 & 3.3 \\
    L$^v_{12}$ & 0.0 & 0.0 & 0.0 & 0.0 & 0.0 & 0.0 & 0.0 & 0.0 & 2.5 & 3.0 & 3.2 & 3.2 & 3.2 \\
    L$^s_{13}$ & 0.0 & 0.0 & 0.0 & 0.0 & 0.0 & 0.0 & 0.0 & 0.0 & 0.0 & 2.0 & 2.4 & 2.5 & 2.6 \\
    L$^v_{13}$ & 0.0 & 0.0 & 0.0 & 0.0 & 0.0 & 0.0 & 0.0 & 0.0 & 2.0 & 2.8 & 3.0 & 3.0 & 3.0 \\
    L$^s_{14}$ & 0.0 & 0.0 & 0.0 & 0.0 & 0.0 & 0.0 & 0.0 & 0.0 & 2.5 & 2.9 & 3.0 & 3.1 & 3.1 \\
    L$^v_{14}$ & 0.0 & 0.0 & 0.0 & 0.0 & 0.0 & 0.0 & 0.0 & 0.0 & 2.5 & 2.9 & 3.0 & 3.1 & 3.1 \\
    L$^s_{15}$ & 0.0 & 0.0 & 0.0 & 0.0 & 0.0 & 0.0 & 0.0 & 0.0 & 1.1 & 2.3 & 2.6 & 2.7 & 2.7 \\
    L$^v_{15}$ & 0.0 & 0.0 & 0.0 & 0.0 & 0.0 & 0.0 & 0.0 & 0.0 & 1.0 & 2.4 & 2.6 & 2.7 & 2.7 \\
    L$^s_{16}$ & 0.0 & 0.0 & 0.0 & 0.0 & 0.0 & 0.0 & 0.0 & 0.0 & 2.8 & 3.3 & 3.4 & 3.4 & 3.2 \\
    L$^v_{16}$ & 0.0 & 0.0 & 0.0 & 0.0 & 0.0 & 0.0 & 0.0 & 0.0 & 2.4 & 3.2 & 3.3 & 3.2 & 3.2 \\
    L$^s_{17}$ & 0.0 & 0.0 & 0.0 & 0.0 & 0.0 & 0.0 & 0.0 & 0.0 & 0.0 & 1.1 & 1.6 & 1.8 & 1.8 \\
    L$^v_{17}$ & 0.0 & 0.0 & 0.0 & 0.0 & 0.0 & 0.0 & 0.0 & 0.0 & 0.0 & 2.1 & 2.5 & 2.6 & 2.7 \\
    L$^s_{18}$ & 0.0 & 0.0 & 0.0 & 0.0 & 0.0 & 0.0 & 0.0 & 0.0 & 2.2 & 2.9 & 3.1 & 3.1 & 3.1 \\
    L$^v_{18}$ & 0.0 & 0.0 & 0.0 & 0.0 & 0.0 & 0.0 & 0.0 & 0.0 & 2.2 & 2.8 & 2.9 & 2.9 & 2.9 \\
    L$^s_{19}$ & 0.0 & 0.0 & 0.0 & 0.0 & 0.0 & 0.0 & 0.0 & 0.0 & 2.2 & 2.9 & 3.0 & 3.0 & 3.1 \\
    L$^v_{19}$ & 0.0 & 0.0 & 0.0 & 0.0 & 0.0 & 0.0 & 0.0 & 0.0 & 2.2 & 2.9 & 3.0 & 3.0 & 3.1 \\
    L$^s_{20}$ & 0.0 & 0.0 & 0.0 & 0.0 & 0.0 & 0.0 & 0.0 & 0.1 & 2.7 & 3.1 & 3.2 & 3.2 & 3.3 \\
    L$^v_{20}$ & 0.0 & 0.0 & 0.0 & 0.0 & 0.0 & 0.0 & 0.0 & 0.1 & 2.7 & 3.0 & 3.3 & 3.2 & 3.3 \\
    \hline
    \hline
    \end{tabular}
    \label{tab:lagrangian_significance}
\end{table*}

\begin{table*}[ht]
    \centering
    \caption{Local significance for $\mathcal{O}_i$ models tested.
    Dashed entries indicate that the observed event is not physical for those WIMP mass and mass splitting combinations.}
    \begin{tabular}{l|l|llllllll}
    \hline
    \multirow{2}{*}{Interaction} & \multirow{2}{*}{Mass (GeV/c$^2$)} & \multicolumn{8}{c}{Mass-splitting, $\delta$ (keV)} \\
                             &      & 0   & 50 & 100  & 150 & 200 & 250 & 300 & 350 \\
    \hline
    \multirow{3}{*}{O$^s_1$} 
    & 400  & 0.0 & 0.0  & 0.8  & 2.2  & 2.6   & 2.9 & 2.9   & -  \\
    & 1000 & 0.0 & 0.0  & 0.8  & 2.2  & 2.7   & 2.9 & 3.0 & 3.3  \\
    & 4000 & 0.0 & 0.0  & 1.0  & 2.3  & 2.7   & 2.9 & 3.0 & 3.3  \\ \hline
    \multirow{3}{*}{O$^v_1$} 
    & 400  & 0.8 & 1.4 & 2.6 & 2.8 & 2.8 & 2.8 & 3.1 & - \\
    & 1000 & 1.3 & 1.6 & 2.6 & 2.8 & 2.9 & 3.0 & 3.4 & 3.4 \\
    & 4000 & 1.1 & 1.7 & 2.6 & 2.9 & 2.9 & 3.1 & 3.4 & 3.4 \\ \hline
    \multirow{3}{*}{O$^s_4$} 
    & 400  & 2.6 & 2.7 & 2.8 & 3.1 & 3.2 & 3.2 & 3.3 & - \\
    & 1000 & 2.7 & 2.8 & 2.8 & 3.0 & 3.2 & 3.2 & 3.4 & 3.4 \\
    & 4000 & 2.8 & 3.0 & 3.0 & 3.1 & 3.2 & 3.3 & 3.4 & 3.4 \\ \hline
    \multirow{3}{*}{O$^v_4$} 
    & 400  & 2.6 & 2.7 & 2.8 & 3.1 & 3.2 & 3.2 & 3.3 & - \\
    & 1000 & 2.7 & 2.8 & 2.8 & 3.0 & 3.2 & 3.2 & 3.3 & 3.4 \\
    & 4000 & 2.8 & 3.0 & 3.0 & 3.1 & 3.2 & 3.3 & 3.4 & 3.4 \\
    \hline
    \hline
    \end{tabular}
    \label{tab:operator_significance}
\end{table*}

\end{document}